\PassOptionsToPackage{table}{xcolor}
\documentclass{article} 
\usepackage{iclr2027_conference,times}

\usepackage{amsmath,amsfonts,bm}

\def\eqref#1{equation~\ref{#1}}

\def\1{\bm{1}}

\newcommand{\valid}{\mathcal{D_{\mathrm{valid}}}}

\DeclareMathAlphabet{\mathsfit}{\encodingdefault}{\sfdefault}{m}{sl}
\SetMathAlphabet{\mathsfit}{bold}{\encodingdefault}{\sfdefault}{bx}{n}

\usepackage{hyperref}
\usepackage{url}
\usepackage[T1]{fontenc}
\usepackage{amsmath,amsfonts,amssymb,amsthm,mathtools,bm}
\usepackage{booktabs}
\usepackage{nicefrac}
\usepackage{microtype}
\usepackage{xcolor}
\usepackage{enumitem}
\usepackage{multirow}
\usepackage{array}
\usepackage{color}
\usepackage[table]{xcolor} 
\usepackage{wrapfig}
\usepackage{multicol}
\usepackage{caption}
\usepackage{diagbox}
\usepackage{pifont}
\usepackage{pdfpages}
\usepackage{arydshln}
\usepackage{makecell}
\usepackage{graphicx}
\usepackage{adjustbox}
\usepackage{colortbl}
\usepackage{siunitx}
\usepackage{pifont}
\usepackage[most]{tcolorbox}
\usepackage{float}
\usepackage{caption} 
\usepackage{tabularx}

\definecolor{blue(pigment)}{rgb}{0.2, 0.2, 0.6}
\definecolor{aliceblue}{rgb}{0.94, 0.97, 1.0}
\definecolor{lightgray}{rgb}{0.88, 0.88, 0.88}
\definecolor{piggypink}{rgb}{0.99, 0.93, 0.97}
\definecolor{mistyrose}{rgb}{1.0, 0.89, 0.88}
\definecolor{deeppurple}{rgb}{0.42, 0.13, 0.42}
\definecolor{lightgray}{rgb}{0.83, 0.83, 0.83}
\definecolor{pastelgray}{rgb}{0.81, 0.81, 0.77}
\definecolor{richmaroon}{rgb}{0.69, 0.19, 0.38}
\definecolor{grey}{rgb}{0.5,0.5,0.5}
\definecolor{vprose}{HTML}{D44973}
\definecolor{vpteal}{HTML}{488CA7}
\definecolor{vpchampagne}{HTML}{CB954D}

\hypersetup{colorlinks=true, linkcolor=richmaroon, citecolor=richmaroon}
\usepackage[capitalise,noabbrev,nameinlink]{cleveref}

\crefformat{figure}{Figure~#2{\color{richmaroon}#1}#3}
\crefformat{table}{Table~#2{\color{richmaroon}#1}#3}
\crefformat{equation}{Equation~#2{\color{richmaroon}#1}#3}
\crefformat{section}{Section~#2{\color{richmaroon}#1}#3}
\crefformat{appendix}{Appendix~#2{\color{richmaroon}#1}#3}
\def\pz{{\phantom{0}}}

\newcommand{\cmark}{\textcolor{teal!70!black}{\ding{51}}}
\newcommand{\xmark}{\textcolor{purple!50!red!80!black}{\ding{55}}}

\newcommand{\hj}[1]{\textcolor{black}{#1}}

\title{\system{}: An Evaluation Benchmark for Language-interfaced Vibe Protein Design}

\author{
{\bfseries Hyunjin Seo}\textsuperscript{1,2},
{\bfseries Hongjoon Ahn}\textsuperscript{1,3},
{\bfseries Jimin Park}\textsuperscript{4},
{\bfseries Sungjun Han}\textsuperscript{1},
{\bfseries Gyubok Lee}\textsuperscript{2},\\
{\bfseries Soojung Yang}\textsuperscript{9, 11},
{\bfseries Cheng-Hao Liu}\textsuperscript{10,11},
{\bfseries Joseph S Brown}\textsuperscript{6},
{\bfseries Leo Chen}\textsuperscript{7},
{\bfseries Gina El Nesr}\textsuperscript{9},\\
{\bfseries Feyisayo Eweje}\textsuperscript{5},
{\bfseries Sarah Gurev}\textsuperscript{7},
{\bfseries Hyejin Lee}\textsuperscript{9},
{\bfseries Junlang Liu}\textsuperscript{8},
{\bfseries Zhihui Qi}\textsuperscript{8},\\
{\bfseries Gyu Rie Lee}\textsuperscript{2},
{\bfseries Sungsoo Ahn}\textsuperscript{2},
{\bfseries Jamin Shin}\textsuperscript{1},
{\bfseries Sangwon Jung}\textsuperscript{1}$^\dag$\\[0.5em]
\textsuperscript{1}Trillion Labs,
\textsuperscript{2}KAIST,
\textsuperscript{3}Seoul National University,
\textsuperscript{4}SK Biopharmaceuticals Co., Ltd.,\\
\textsuperscript{5}MIT,
\textsuperscript{6}University of Toronto,
\textsuperscript{7}Harvard Medical School,
\textsuperscript{8}Harvard University,\\
\textsuperscript{9}Stanford University,
\textsuperscript{10}California Institute of Technology,
\textsuperscript{11}FutureHouse
}

\iclrfinalcopy 
\begin{document}

\AddToShipoutPictureFG*{
    \put(105,760){
        \includegraphics[width=3.2cm]{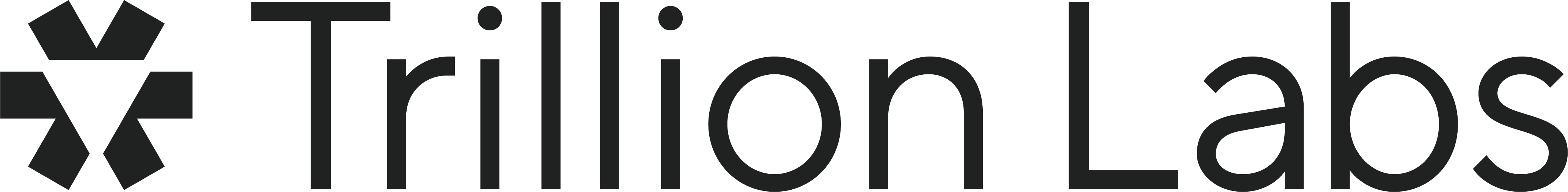}
    }
}
\maketitle

\def\thefootnote{$\dag$}\footnotetext{Corresponding Author.}

\begin{abstract}
Protein design aims to compose amino-acid sequences that fold into stable three-dimensional structures while satisfying targeted functional properties. The field is increasingly shifting toward \emph{vibe protein design}, where a single model is expected to generate novel sequences, engineer existing proteins, and reason about protein characteristics through flexible natural-language constraints. 
However, existing benchmarks typically evaluate isolated aspects of protein design or assume predefined structured inputs, making them ill-suited to assess this broad, open-ended setting. 
To address this gap, we present Vibe Protein Design Benchmark (\system{}), a language-interfaced benchmark that evaluates whether a single model can operate across three complementary stages of a computational protein-design workflow: \textbf{\emph{recognition}}, \textbf{\emph{engineering}}, and \textbf{\emph{generation}}. Each stage is grounded in expert-curated mechanistic rationales and multi-faceted in silico validation, to computationally verify whether model outputs are biologically plausible. 
Across diverse general-purpose and domain-specialized LLMs, evaluated with and without tool access, no model performs strongly across all three stages at once.
Beyond measuring end-task success, \system{} exposes where current systems break: we identify recurring failure patterns and characterize how tool-augmented agents select tool arguments. Together, these findings show that vibe protein design remains a substantial open challenge.
\end{abstract}

\section{Introduction}\label{01_introduction}

Protein design aims to compose amino-acid sequences that fold into stable three-dimensional structures while satisfying targeted functional properties. Owing to its broad applicability across therapeutic development, agriculture, and biotechnology, it has attracted significant attention as a promising direction for computational modeling and scientific discovery~\citep{proteindesign_modelcentric, adaptyv_egfr, notin2024machine, nos}. This interest has driven rapid progress in deep learning approaches to computational protein design, advancing tasks such as protein engineering~\citep{hie2024efficient, kirjner2023improving, hie2022adaptive, thermompnn} and de novo sequence generation~\citep{boltzgen, proteina-complexa, bindcraft}. However, these approaches have largely been developed as task-specific models that consume \emph{predefined, structured} inputs such as backbone coordinates, numerical descriptors, or other functional constraints~\citep{boltzgen, proteina-complexa, bindcraft, protgpt2}. Such formulations are effective in controlled settings, but offer limited flexibility for open-ended objectives, where practitioners may express more naturally at a high level such as designing an enzyme capable of degrading PFAS.

Recent work has begun to explore models that take flexible biological objectives expressed in natural language, infer the corresponding task requirements from relevant protein context, and carry out the appropriate protein-design task~\citep{proteindt, proteincrow, idr_llmagent, scireasoner}. We refer to this setting as \textit{vibe protein design}, drawing an analogy to \textit{vibe coding} in software development~\citep{gu2026matthew, li2026vibe, meske2025vibe}. 

These objectives, however, are rarely satisfied by a single task. \emph{Generating} a protein entails \emph{recognizing} its biophysical and functional properties to validate its quality~\citep{He2026ORI, Qing2022ProteinDesign}, and the resulting sequence often requires further \emph{engineering} to enhance a targeted property~\citep{sappington2026improved, cho2025stable}.
Large language models (LLMs) have increasingly been investigated in this role, as their language interface admits biological intents of diverse tasks and their reasoning enables orchestrating the tools each demands~\citep{gpt4b_micro, gpt-rosalind, naturelm, prollama, protllm, Proteingpt, swarms, proteincrow, idr_llmagent, toursynbio, scireasoner, evolla, prottex}. Nevertheless, whether one model can perform reliably across all three capabilities, with or without access to external tools, remains unclear.


As this direction gains momentum, rigorously evaluating these models becomes increasingly essential. A benchmark for vibe protein design should therefore assess whether a single model can operate across this connected workflow under natural-language design intents. Existing benchmarks, however, only partially satisfy this requirement, either focusing on isolated aspects of protein design~\citep{pdfbench, liveproteinbench} or relying on predefined structured inputs~\citep{Pdb-struct, proteinbench}, and therefore fall short of an integrated framework for language-interfaced settings.

\begin{figure}[t]
    \centering
    \includegraphics[width=\linewidth]{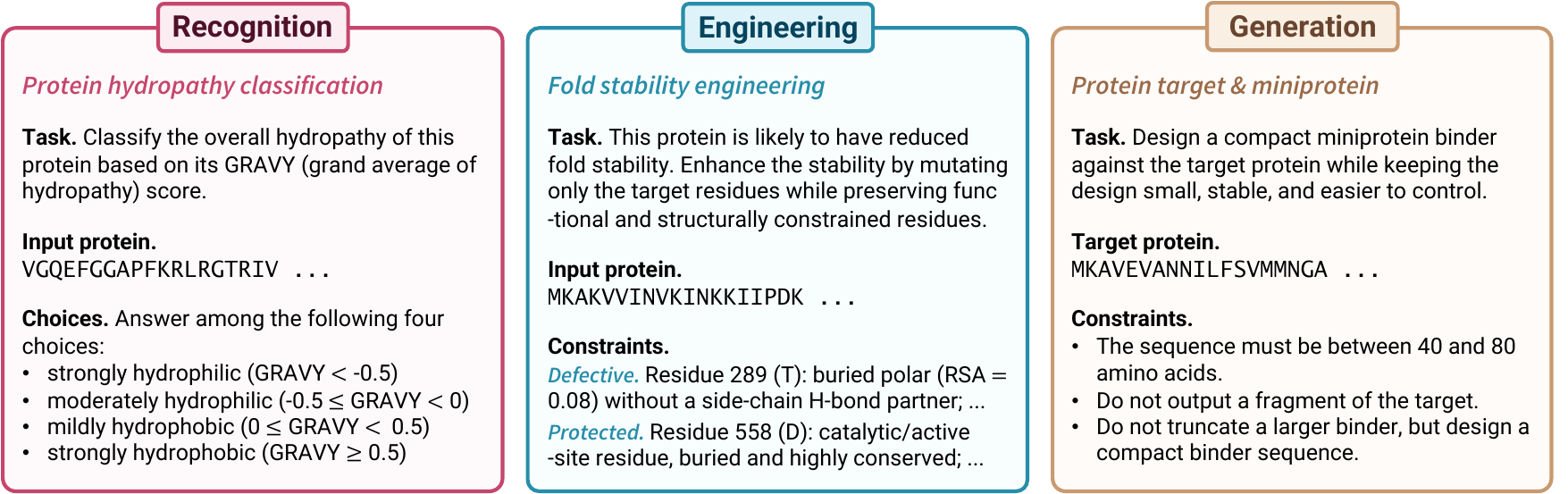}
    \caption{
    Representative queries from the three stages of \system{}. Each query supplies a natural-language task, the protein or target context, and the stage-specific constraints.} 
\label{fig:query-examples}
\vskip -10pt
\end{figure}

To address this gap, we present Vibe Protein Design Benchmark (\system{}), a language-interfaced benchmark that operationalizes this workflow as three evaluation stages: \textbf{\emph{recognition}}, \textbf{\emph{engineering}}, and \textbf{\emph{generation}}. \system{} instantiates each stage through natural-language queries paired with the protein context needed to express realistic design intents. Recognition uses question-answering over sequence, structural, and functional properties; engineering asks models to edit existing proteins using expert-curated mechanistic rationales; and generation evaluates novel protein design from functional or target-grounded specifications. Every stage carries a computationally verifiable evaluation protocol built on strong evaluators~\citep{protenix-v1, cho2026accurate, boltz2}, from exact-answer assessment to multi-faceted in silico validation. We validate queries and evaluators through human expert review and in silico evaluators comparison (Appendix~\ref{B}).

We evaluate general-purpose LLMs~\citep{gpt-5, opus4.6, gemini-3.1, deepseekv3.2, deepseekv4, qwen3.5, kimi-k2.5} on \system{} in both tool-free and tool-augmented agent settings, and additionally evaluate domain-specialized LLMs~\citep{naturelm, scireasoner, txgemma} to assess whether scientific pretraining confers an advantage. Our results reveal that \textit{no single model performs strongly across all three stages at once}, with pass rates dropping sharply from recognition to engineering and generation. Beyond measuring end-task success, \system{} exposes where current systems break: we identify recurring failure patterns and characterize how tool-augmented agents select and misuse tool arguments. We further find that scores are correlated across biologically linked subtasks spanning different stages, suggesting that \system{} captures coherent protein-design competence rather than a collection of isolated probes. Together, these findings show that vibe protein design remains a substantial open challenge while validating the integrated structure of the benchmark.


\section{Related Work}\label{02_related_work}

\noindent\textbf{Non-language-based protein design benchmarks.}
Early protein design benchmarks primarily evaluate models under structured input schemas, such as backbone coordinates, geometric constraints, motif residues, or predefined target properties. These benchmarks have been central to measuring core design capabilities such as inverse folding, foldability, structural recovery, and mutation-effect prediction~\citep{Pdb-struct, proteininvbench, proteinbench, venusfactory}. However, because their inputs are not expressed in language, they are limited in evaluating open-ended biological intents that practitioners may describe flexibly in text.

\noindent\textbf{Language-interfaced protein design benchmarks.}
Recent benchmarks have begun to evaluate protein design through natural-language interfaces. One line of work evaluates sequence generation conditioned on a natural-language function description or a bound ligand~\citep{pdfbench, instructpro, Mol-instructions}, but limits its evaluation to a subset of generation tasks. A parallel line evaluates LLM agents that orchestrate external protein-design tools, scoring how they plan a design workflow, execute the appropriate software, and interpret the resulting validation outputs~\citep{biodesignbench, able, venusfactory2}, yet reports performance only in the tool-augmented setting, leaving the internal design competence of LLMs unmeasured. In contrast, \system{} queries a broad spectrum of the protein-design workflow, spanning recognition, engineering, and generation, and evaluates LLMs under both tool-free and tool-augmented settings. Its generation stage covers both semantic functional specifications and explicit binding targets. We provide a detailed comparison with more related literature in Appendix~\ref{F}.

\section{Task Formulation}\label{03_task_formulation}

We formulate vibe protein design as a setting in which a practitioner states an open-ended biological objective in natural language with the protein or target context, and the model must resolve it into a concrete, task-specific response. Conventional formulations instead assume a predefined input schema such as a numerical target or a geometric constraint~\citep{Pdb-struct, proteinbench, peer, care}, which fixes both the form and the granularity of what can be stated. A language interface fixes neither, carrying a coarse functional goal and a mechanistic rationale in the prompt: a practitioner may ask whether an uncharacterized sequence shares a homolog's fold despite low sequence identity, how to improve the thermostability of an enzyme while preserving catalytic activity, or how to design an enzyme for a reaction with no natural counterpart. 

Notably, these differing outputs correspond to distinct but interdependent design capabilities, spanning a broad workflow of understanding, refinement, and creation. Verifying a generated protein requires recognizing its biophysical and functional characteristics~\citep{He2026ORI, Qing2022ProteinDesign}, and the protein itself frequently has to be engineered to satisfy the intended properties~\citep{sappington2026improved, cho2025stable}. Such engineering decisions, in turn, depend on identifying which sequence or structural features to change and which to preserve~\citep{Goldenzweig2016PROSS}, a recognition problem in its own right. We accordingly organize \system{} around three complementary stages that together cover this workflow: protein recognition, rationale-guided engineering, and functional protein generation. 

\noindent\textbf{Protein recognition} evaluates whether a model can infer biologically meaningful properties from a protein sequence. This stage probes understanding across the sequence--structure--function paradigm, which grounds downstream protein design decisions~\citep{sandhya2016protein, hamamsy2023learning}. 

\noindent\textbf{Rationale-guided engineering} evaluates whether a model can internalize mechanistic rationales and translate them into sequence modifications that enhance a target property. In practice, human experts identify residues responsible for a design deficiency ~\citep{chubb2026rational} while preserving residues critical to fold and function~\citep{teufl2022engineering}. We capture this workflow by instructing the model to repair defective target positions and keep protected positions intact.

\noindent\textbf{Functional protein generation} evaluates whether a model can generate novel protein sequences from functional design intent. This stage covers diverse generation scenarios, ranging from semantic biological functions that are difficult to express as a rigid numerical or geometric constraint~\citep{scireasoner, ProCALM, pinal} to target-grounded binder design where the specific binding target serves as the design constraint~\citep{boltzgen, proteina-complexa, bindcraft}. 
\section{Benchmark Design and Construction}\label{04_benchmark_construction}

\begin{figure}[t!]
    \centering
    \includegraphics[width=1.0\linewidth]{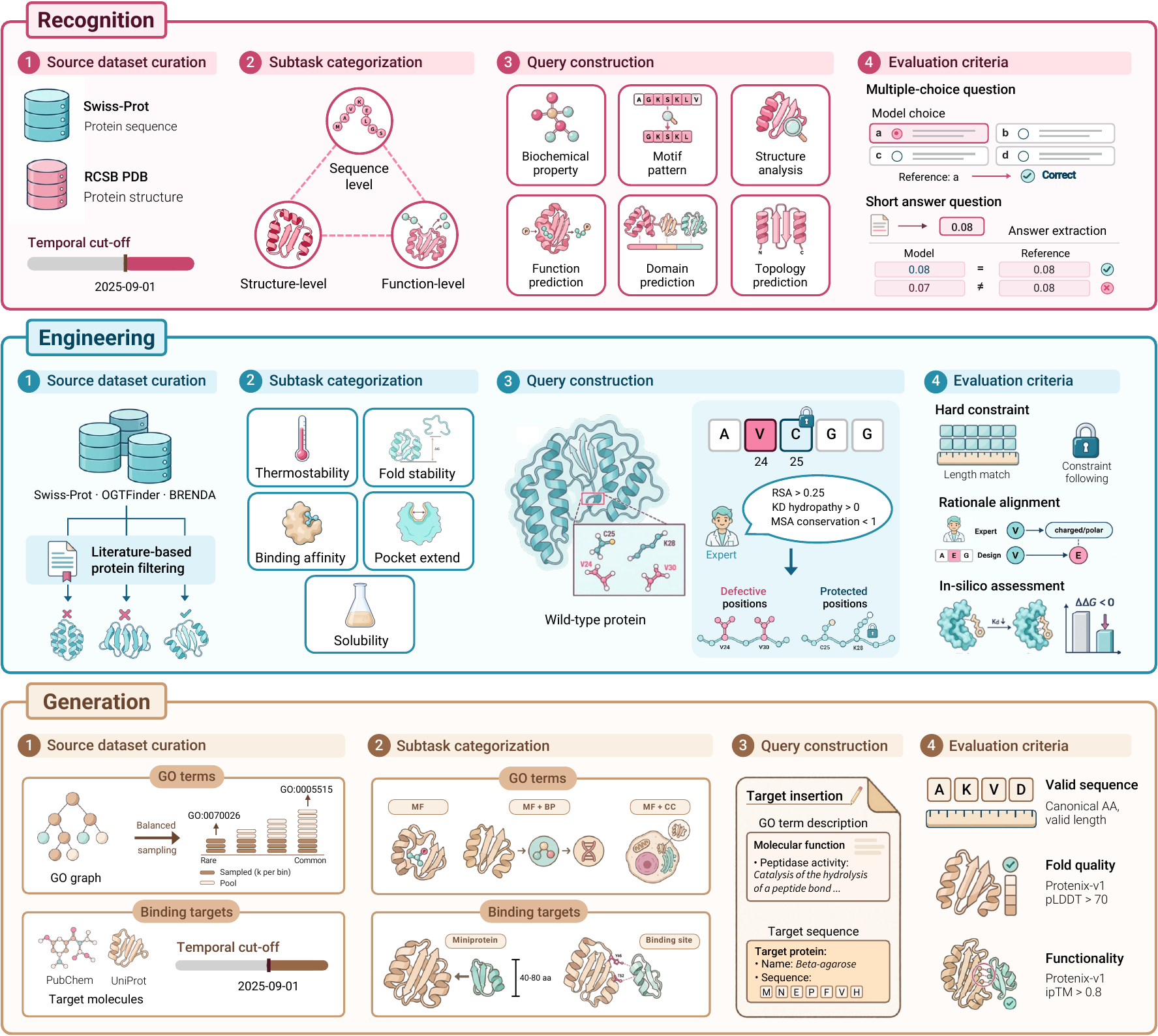}
    \caption{
    Evaluation dataset construction pipeline of \system{}. 
\system{} consists of three stages mirroring a computational protein design workflow: recognition (top), engineering (middle), and generation (bottom). 
Each stage is built through a shared four-step process: \textbf{source dataset curation} collects proteins, functional terms, or molecular targets from public repositories with stage-specific contamination control, \textbf{subtask categorization} defines target design intents, \textbf{query construction} illustrates constraints included in queries, and \textbf{evaluation criteria} assess model outputs through exact match or multi-faceted in silico evaluators.
    }
    \label{fig:benchmark-construction}
    \vskip -10pt
\end{figure}

In this section, we describe the construction of \system{} across its three primary stages: protein recognition (\cref{4.1}), rationale-guided protein engineering (\cref{4.2}), and functional protein generation (\cref{4.3}). Our construction strategy is designed to make each stage cover representative and practically relevant scenarios that arise in language-interfaced protein design. 
To this end, we first consult domain experts to identify stage-specific subtasks that
reflect practical expert intents and core design capabilities. We then curate the proteins,
functional terms, or molecular targets needed to instantiate them from public
repositories under a strict temporal cutoff that reduces contamination from the training
corpora of contemporary LLMs. Finally, we establish stage-specific evaluation rubrics that combine hard constraints with multi-faceted in silico validation. \cref{fig:query-examples} illustrates representative queries and \cref{fig:benchmark-construction} summarizes the construction pipeline.

\subsection{Protein recognition}\label{4.1}

\noindent\textbf{Subtask categorization.}  The recognition stage is grounded in the canonical sequence--structure--function paradigm~\citep{sandhya2016protein, koehler2023sequence}, which links protein sequences to three-dimensional structures and biological functions. We therefore organize subtasks into three corresponding levels. 
The sequence-level probes local and global physicochemical properties such as motif detection and net charge prediction; the structure-level evaluates secondary structure, residue burial, and disulfide bonding; and the function-level assesses evolutionary families and functional roles. These layers identify which positions are modifiable liabilities, which must be preserved, and which functional sites can be carried over from related families in downstream protein design.

\noindent\textbf{Source dataset curation.}  We curate proteins from the UniProtKB/Swiss-Prot database~\citep{uniprot} whose experimentally resolved 3D structures are deposited in the RCSB PDB database~\citep{RCSB_PDB}. To minimize contamination risk from the training corpora of contemporary LLMs, we restrict the dataset to UniProt entries registered after September 1, 2025.

\noindent\textbf{Query construction.} Each query pairs a protein sequence with a natural-language question that names the property to be identified, phrased as a practitioner would pose it during analysis. We derive the reference property values from computational tools and curated databases, including Biopython descriptors~\citep{biopython}, DSSP assignments~\citep{dssp} on experimentally resolved structures, and domain and function annotations from ECOD~\citep{ecod}, InterPro~\citep{interpro}, and Gene Ontology (GO)~\citep{go_benchmark}. 
Although these labels derive from experimental structures and curated annotations, each query exposes only the amino-acid sequence, so the model must recover the property from its own knowledge or by selecting an appropriate tool to analyze or fold the sequence. We pose categorical properties as multiple-choice questions, and elicit deterministic structured outputs such as residue-level attributes or functional position identifiers through short-answer generation.

\noindent\textbf{Evaluation.} We evaluate each query according to its format. For multiple-choice, a response is correct when the model selects the ground-truth option. For short-answer generation, we use an exact match against the deterministic reference answer after output normalization. 

\subsection{Rationale-guided engineering}\label{4.2}

\noindent\textbf{Subtask categorization.}  The engineering stage is designed to reflect property-driven protein optimization, where practitioners aim to improve a specific biochemical property while preserving structural and functional integrity. We organize subtasks around three objectives that span the primary dimensions along which proteins are routinely engineered in practice~\citep{korendovych2017rational, grigorakis2025protein}: solubility, stability (including fold stability and thermostability), and activity (including binding affinity and pocket expansion). 

\noindent\textbf{Source dataset curation.} 
We draw wild-type proteins from the UniProt/Swiss-Prot database for solubility and fold stability, the OGTFinder database~\citep{ogtfinder} for thermostability, and the BRENDA enzyme database~\citep{brenda} for activity. To reduce contamination from previously engineered targets, we apply a rule-based literature filter before constructing editing queries. We query PubMed and Semantic Scholar for each wild type using task-specific engineering search terms, and discard it when a returned title or abstract metadata simultaneously contains the protein name, a protein-design keyword (\eg, rational protein design), and at least one task-context keyword (\eg, stability engineering). We detail the full search terms in Appendix~\ref{D}.

\noindent\textbf{Expert-curated design rationales.} 
Protein engineering in practice is guided by mechanistic rationales that identify which residues carry a specific deficiency and in which direction to repair them, while leaving intact the residues critical to fold and function. To capture this process, we formalize each engineering task as a rationale-guided mutation design problem, in which the model must mutate a designated set of defective positions while leaving protected positions unchanged. Since each engineering objective recurs with similar mechanistic logic, we elicit rationale templates per subtask from domain experts. Each template names the liability to be repaired, the admissible mutation directions for the target property, the residue classes that must be preserved, and the concrete criteria by which each of these is recognized. 
We then instantiate each template on every wild-type protein by computing the features its criteria refer to, combining structural and biochemical analysis with Henikoff-weighted MSA conservation statistics~\citep{henikoff1994position}. Applying the criteria to these features yields the disjoint defective and protected sets.
The complete set of rationale templates and criteria is provided in Appendix~\ref{D}.

\noindent\textbf{Query construction.}  Each query presents a wild-type sequence, a target property to optimize, and the rationale-derived defective and protected residues. The model is tasked with proposing a mutant that improves the property by \hj{engineering} only the defective positions while leaving protected positions unchanged. This formulation evaluates whether the model can \emph{internalize and apply mechanistic design rationales}, rather than relying on unconstrained sequence generation. To ensure that the provided rationales are actionable, we validate each query by substituting defective positions with admissible reference residues and retaining only those cases where the substitutions yield measurable improvements under our in silico evaluation pipeline. We additionally construct a constraint-free variant of each query that omits the defective and protected sets and supplies only the wild-type and the target property, requiring the model to localize the liability and choose the repair on its own. 

\noindent\textbf{Evaluation.}  We evaluate each proposed mutant using three criteria. Hard constraints (Hard const.) verify whether the mutant preserves the wild-type length, keeps protected positions unchanged, performs at least one mutation, and \hj{mutates} only defective targets. Rationale alignment (Rat align.) checks whether each mutation follows the expert-curated repair direction for the target property. In silico validation (In silico valid.) assesses biological plausibility using confidence from co-folding models~\citep{protenix-v1, boltz2} and property-specific validators~\citep{cho2026accurate, boltz2, pykvfinder}. We omit rationale alignment for constraint-free queries, since a repair direction cannot be assessed automatically once a model mutates arbitrary positions. We provide the full rationale templates, literature search terms, and evaluator thresholds in Appendix~\ref{D}.

\subsection{Functional protein generation}\label{4.3}

\noindent\textbf{Subtask categorization.}  The generation stage covers two ways practitioners express open-ended design intent: semantic functional descriptions and explicit molecular targets. The first track addresses functional roles that are difficult to formalize as structured constraints, making natural language an ideal way to convey them~\citep{pdfbench, Mol-instructions, prollama, ProCALM, pinal}. We draw these specifications from language-based functional definitions of Gene Ontology (GO) terms, ranging from a single molecular function (MF) to MF paired with biological process (BP) or cellular component (CC). The resulting terms span broad protein classes from enzymatic catalysis and nucleic-acid processing to transcription factor, transporter, and motor activity. The second track focuses on binder design for specific molecular targets, where the target representation itself (\eg, SMILES or protein sequence) serves as the design specification~\citep{proteincrow, proteinbench, instructpro, boltzgen, bindcraft}. We further divide this track into small-molecule and protein-target settings, and the protein-target task additionally carries two variants drawn from practical laboratory demands: a miniprotein length constraint for controllability~\citep{baker2017miniprotein, ozga2022design}, and a binding-site constraint naming the residues the binder must contact, reflecting therapeutic applications that depend on engaging a particular receptor surface~\citep{rfdiffusion_antibody}. Beyond these tasks, we report a supplementary set of antibody, DNA, and RNA binder design in \cref{tab:additional-binder-generation}.

\noindent\textbf{Source dataset curation.} For GO-conditioned generation, we sample single terms and co-occurring term pairs across information-content (IC) bins~\citep{go_benchmark}, using marginal and joint annotation frequencies, respectively, so that queries are not dominated by overly common annotations (Appendix~\ref{E}). For binder generation, we apply a strict temporal holdout (2025-09-01) when curating novel protein targets from UniProt and small-molecule targets from PubChem.

\noindent\textbf{Query construction.}  
Each query provides either a natural-language functional specification or a molecular target representation, and asks the model to generate a novel protein sequence conditioned on that objective. The miniprotein subtask additionally requires a sequence of 40 to 80 residues, and the binding-site subtask enumerates the target residues the binder must contact, each written as a one-letter amino acid followed by its one-indexed position in the target sequence (\eg, Y46).

\noindent\textbf{Evaluation.}  
We evaluate generated sequences along three axes. Sequence validity (Seq. valid.) checks canonical amino-acid composition and, where applicable, the miniprotein length constraint. Fold plausibility (Fold valid.) is assessed from Protenix-v1~\citep{protenix-v1} confidence. Functional consistency (Func. Cons.) uses a different measure for each track. For GO-conditioned generation, we measure functional consistency with an ESM-C-based~\citep{esmfold2} GO-reference matching rule that accepts a sequence when its minimum cosine similarity to ten reference proteins sharing the designated GO terms is at least the minimum among those references. For binder generation, we use ipTM from Protenix-v1, following established practice~\citep{pxdesign, halludesign}, and the binding-site subtask further requires the binder to contact at least half of the specified site residues within a 5~\AA{} heavy-atom cutoff. We additionally report novelty against natural proteins, and the provide full set of validators and thresholds in Appendix~\ref{E}.

\subsection{Quality control}\label{4.4}
We apply a two-stage expert review to ensure that each subtask and query in \system{} reflects a meaningful protein design competence. In the primary consultation, domain experts select the subtasks that define the scope of each stage. In a secondary pass, a separate group of experts who did not participate in the primary consultation scores each query with its evolutionary and structural context, the tools used to construct it, and its answer evaluation method. For each query, experts give a yes/no judgment against 5-6 rubric items spanning realism/meaningfulness, clarity, and in silico verifiability. We retain subtasks whose mean score across evaluators is at least 0.5, ranging from 0 (unanimous rejection) to 1 (unanimous endorsement). We provide the full rubric in Appendix~\ref{G}.

\section{Experiments}\label{05_experiments}

\subsection{Experimental setup}\label{5.1}
We evaluate eight general-purpose LLMs on \system{} under two settings: a tool-free setting that isolates their internal protein-design competence, and a tool-augmented agent setting based on ReAct~\citep{react} whose tool list we detail in Appendix~\ref{A}. The general-purpose group comprises GPT-5.4~\citep{gpt-5}, Gemini3.1-Pro~\citep{gemini-3.1}, Opus-4.6~\citep{opus4.6}, DeepSeek-V4-Pro~\citep{deepseekv4}, DeepSeek-V3.2~\citep{deepseekv3.2}, Kimi-K2.5~\citep{kimi-k2.5}, and Qwen3.5~\citep{qwen3.5} at 397B-A17B and 9B scales. We further evaluate a domain-specialized group: NatureLM-8$\times$7B~\citep{naturelm}, SciReasoner-8B~\citep{scireasoner}, and TxGemma-chat~\citep{txgemma} at 9B and 27B scales. We additionally include five non-LLM reference baselines, the protein-text multimodal models ProteinDT~\citep{proteindt}, ProDVa~\citep{ProDVa}, and PAAG~\citep{paag}, and the binder design models BoltzGen~\citep{boltzgen} and BindCraft~\citep{bindcraft}. We report pass rate as the primary measure, counting a query as passed only when the output satisfies every criterion defined for its subtask in \cref{04_benchmark_construction}, and counting a query declined by internal safety filters as a failure. Appendix~\ref{B} reports further analyses, including error-level agent comparisons, unbounded iteration under an advanced setup, tool-argument choices, a temporal-cutoff ablation, and antibody, RNA, and DNA binder generation.

\begin{table}[t]
\centering\Huge
\caption{
Pass rates (\%) for each evaluation criterion across the three stages of \system{}, reported for general-purpose and domain-specialized LLMs without tools. Pass-rate columns are shaded, and within each, the best and second-best results are shown in \textbf{bold} and \underline{underline}.
}
\label{tab:main-rubric}
\resizebox{\textwidth}{!}{
\begin{tabular}{l >{\columncolor{piggypink!50}}c c c c >{\columncolor{piggypink!50}}c c c c c >{\columncolor{piggypink!50}}c}
\toprule
& \multicolumn{1}{c}{Recognition}
& \multicolumn{4}{c}{Engineering}
& \multicolumn{5}{c}{Generation}\\
\cmidrule(lr){2-2}
\cmidrule(lr){3-6}
\cmidrule(lr){7-11}
Models
& Pass rate
& Hard const.
& Rat. align.
& In silico valid.
& Pass rate
& Seq. valid.
& Fold valid.
& Func. Cons.
& Novelty
& Pass rate \\
\midrule
\rowcolor{grey!3}
\emph{General-purpose}
& \cellcolor{piggypink!50}
& & & & \cellcolor{piggypink!50}
& & & & & \cellcolor{piggypink!50} \\
GPT-5.4 & \textbf{82.6} & \underline{86.7} & \textbf{76.7} & \textbf{56.7} & \textbf{56.7} & 95.7 & 54.8 & \underline{18.6} & 67.2 & 7.6 \\
Gemini3.1-Pro & 56.0 & 63.3 & 51.7 & 38.3 & 38.3 & \underline{99.0} & \textbf{74.8} & 18.1 & 61.2 & \underline{14.8} \\
Opus-4.6 & \underline{75.4} & 5.0 & 5.0 & 3.3 & 3.3 & \textbf{99.5} & 38.6 & 13.3 & 81.7 & 2.9 \\
DeepSeek-V4-Pro & {73.8} & 66.7 & 48.3 & 40.0 & 40.0 & \textbf{99.5} & \underline{65.7} & 17.6 & 70.9 & 11.4 \\
DeepSeek-V3.2 & 70.7 & 65.0 & 55.0 & 40.0 & 40.0 & 98.6 & 61.4 & 11.4 & 79.9 & 5.7 \\
Kimi-K2.5 & 58.4 & 6.7 & 3.3 & 1.7 & 1.7 & 94.8 & 39.0 & 10.5 & 87.1 & 4.8 \\
Qwen3.5-397B-A17B & 72.1 & 63.3 & \underline{58.3} & \underline{45.0} & \underline{45.0} & \textbf{99.5} & 56.7 & 12.4 & 89.3 & 5.7 \\
Qwen3.5-9B & 40.8 & 1.7 & 0.0 & 0.0 & 0.0 & 82.9 & 43.3 & 6.2 & 88.8 & 2.9 \\
\arrayrulecolor{black!40}\midrule
\arrayrulecolor{black}
\rowcolor{grey!3}
\emph{Domain-specialized}
& \cellcolor{piggypink!50}
& & & & \cellcolor{piggypink!50}
& & & & & \cellcolor{piggypink!50} \\
NatureLM-8$\times$7B & 26.5 & 73.3 & 3.3 & 3.3 & 3.3 & 96.7 & 55.2 & 16.7 & 78.8 & \textbf{15.7} \\
SciReasoner-8B & 40.5 & \textbf{91.7} & 5.0 & 3.3 & 3.3 & 98.6 & 46.2 & \textbf{19.5} & 74.6 & 13.8 \\
TxGemma-9B & 40.8 & 5.0 & 0.0 & 0.0 & 0.0 & \textbf{99.5} & 49.0 & 8.6 & \underline{92.0} & 3.8 \\
TxGemma-27B & 33.7 & 13.3 & 0.0 & 0.0 & 0.0 & 93.3 & 37.1 & 5.2 & \textbf{96.5} & 1.0 \\
\bottomrule
\end{tabular}
}
\vskip -5pt
\end{table}
\begin{table}[t]
\centering
\Huge
\caption{
Pass rates for each evaluation criterion across the three stages of \system{} with and without tool use. The best performance is represented by bold.
}
\label{tab:tool-rubric}
\setlength{\tabcolsep}{4pt}
\resizebox{\textwidth}{!}{
\begin{tabular}{l l >{\columncolor{piggypink!50}}c c c c >{\columncolor{piggypink!50}}c c c c c >{\columncolor{piggypink!50}}c}
\toprule
& & \multicolumn{1}{c}{Recognition}
& \multicolumn{4}{c}{Engineering}
& \multicolumn{5}{c}{Generation} \\
\cmidrule(lr){3-3}
\cmidrule(lr){4-7}
\cmidrule(lr){8-12}
Models & Setup
& Pass rate
& Hard const. & Rat. align. & In silico valid. & Pass rate
& Seq. valid. & Fold valid. & Func. Cons. & Novelty & Pass rate \\
\midrule
\multirow{2}{*}{GPT-5.4}
& w/o tool & 82.6 & 86.7 & 76.7 & 56.7 & 56.7 & 95.7 & 54.8 & 18.6 & 67.2 & 7.6 \\
& w/ tool & \textbf{84.7} & 95.0 & 75.0 & 60.0 & \textbf{60.0} & 97.1 & 57.6 & 23.8 & 82.5 & \textbf{12.4} \\
\arrayrulecolor{black!40}\midrule
\arrayrulecolor{black}
\multirow{2}{*}{Gemini3.1-Pro}
& w/o tool & 56.0 & 63.3 & 51.7 & 38.3 & 38.3 & 99.0 & 74.8 & 18.1 & 61.2 & 14.8 \\
& w/ tool & \textbf{82.3} & 68.3 & 60.0 & 45.0 & \textbf{45.0} & 75.2 & 65.2 & 22.9 & 70.1 & \textbf{20.0} \\
\arrayrulecolor{black!40}\midrule
\arrayrulecolor{black}
\multirow{2}{*}{DeepSeek-V4-Pro}
& w/o tool & 73.8 & 66.7 & 48.3 & 40.0 & 40.0 & 99.5 & 65.7 & 17.6 & 70.9 & 11.4 \\
& w/ tool & \textbf{84.8} & 86.7 & 75.0 & 55.0 & \textbf{55.0} & 94.8 & 76.7 & 28.6 & 78.9 & \textbf{23.8} \\
\arrayrulecolor{black!40}\midrule
\arrayrulecolor{black}
\multirow{2}{*}{DeepSeek-V3.2}
& w/o tool & 70.7 & 65.0 & 55.0 & 40.0 & 40.0 & 98.6 & 61.4 & 11.4 & 79.9 & 5.7 \\
& w/ tool & \textbf{74.0} & 73.3 & 65.0 & 56.7 & \textbf{56.7} & 59.5 & 45.2 & 14.3 & 88.2 & \textbf{12.4} \\
\arrayrulecolor{black!40}\midrule
\arrayrulecolor{black}
\multirow{2}{*}{Qwen3.5-397B-A17B}
& w/o tool & 72.1 & 63.3 & 58.3 & 45.0 & 45.0 & 99.5 & 56.7 & 12.4 & 89.3 & 5.7 \\
& w/ tool & \textbf{76.2} & 88.3 & 80.0 & 65.0 & \textbf{65.0} & 63.3 & 43.3 & 8.6 & 84.2 & \textbf{6.2} \\
\bottomrule
\end{tabular}
}
\vskip -10pt
\end{table}

\begin{table}[t]
\centering\Huge
\caption{Pass rates for each evaluation criterion on engineering and generation tasks for non-LLM baselines. BindCraft is evaluated only on protein-target binder generation queries, whereas BoltzGen is evaluated on all binder generation queries.}
\label{tab:non-llm-multimodal-rubric-summary}
\resizebox{\textwidth}{!}{
\begin{tabular}{l c c c >{\columncolor{piggypink!50}}c c c c c >{\columncolor{piggypink!50}}c}
\toprule
 & \multicolumn{4}{c}{Engineering} & \multicolumn{5}{c}{Generation}\\
\cmidrule(lr){2-5} \cmidrule(lr){6-10}
Models & Hard const. & Rat. align. & In silico valid. & Pass rate & Seq. valid. & Fold valid. & Func. Cons. & Novelty & Pass Rate \\
\midrule
ProteinDT & 0.0 & 0.0 & 0.0 & 0.0 & 100.0 & 30.5 & 4.8 & 94.2 & 0.0 \\
PAAG & 0.0 & 0.0 & 0.0 & 0.0 & 100.0 & 24.3 & 2.9 & 95.6 & 0.5 \\
ProDVa & 0.0 & 0.0 & 0.0 & 0.0 & 82.9 & 65.2 & 31.0 & 43.1 & 19.0 \\
BindCraft & N/A & N/A & N/A & N/A & 98.9 & 91.1 & 21.1 & 96.5 & 20.0 \\
BoltzGen & N/A & N/A & N/A & N/A & 100.0 & 95.8 & 36.7 & 69.7 & 36.7 \\
\arrayrulecolor{black}\bottomrule
\end{tabular}
}
\vskip -10pt
\end{table}

\subsection{Main results}\label{5.2}
\cref{tab:main-rubric} reports all LLM baselines without tools. Overall, \textbf{no model achieves strong performance across all three stages}: the best pass rate reaches 82.6 in recognition but drops to 56.7 in engineering and 15.7 in generation, leaving a substantial portion of engineering and generation queries unsolved even for the strongest models. These failures share a common shape, in which a model satisfies the surface requirements of a query while missing its substantive one. In engineering, hard-constraint compliance does not entail adherence to the mechanistic rationale: SciReasoner-8B and NatureLM-8$\times$7B reach 91.7 and 73.3 on hard constraints, yet their rationale alignment falls to 5.0 and 3.3. Opus-4.6 fails for two distinct reasons, as its content filter declines 35.0\% of its engineering queries and its edits land outside the permitted set in 36.7\% of cases. Generation exhibits the same dissociation in a more extreme form. Nearly all models produce valid sequences, yet functional consistency stays below 20 for every model. NatureLM-8$\times$7B and SciReasoner-8B lead the generation pass rate, indicating that pretraining on protein data does transfer, but their functional consistency of 16.7 and 19.5 stays in the similar range as the general-purpose models. Together, these results identify mechanistic and functional grounding as a key bottleneck for current models. 

\cref{tab:tool-rubric} further compares the five strongest models with their tool-augmented counterparts. Tool access raises every model-by-stage pass rate, to 84.8 in recognition and 23.8 in generation for DeepSeek-V4-Pro and to 65.0 in engineering for Qwen3.5-397B-A17B. Nevertheless, a substantial fraction of engineering and generation queries remains unsolved, with the best models still failing 35.0\% and 76.2\%, respectively. Tool access therefore complements but does not substitute for the design competence \system{} probes.

\subsection{Quantitative analysis}\label{5.3}

\noindent
\begin{wraptable}{r}{0.45\linewidth}
\vspace{-\baselineskip}
\centering\small
\captionof{table}{
Pass rates for constraint-free engineering in tool-augmented setting.
}
\label{tab:constraint-free-engineering}
\renewcommand{\arraystretch}{0.85}
\resizebox{\linewidth}{!}{
\begin{tabular}{l c c >{\columncolor{piggypink!50}}c}
\toprule
Models
& \makecell{Hard\\const.}
& \makecell{In silico\\valid.}
& \makecell{Pass\\rate} \\
\midrule
GPT-5.4 & 96.7 & 55.0 & 55.0 \\
Gemini3.1-Pro & 56.7 & 30.0 & 30.0 \\
DeepSeek-V4-Pro & 75.0 & 55.0 & 55.0 \\
DeepSeek-V3.2 & 46.7 & 30.0 & 30.0 \\
Qwen3.5-397B-A17B & 75.0 & 38.3 & 38.3 \\
\bottomrule
\end{tabular}
}
\vskip -10pt
\end{wraptable}

\noindent\textbf{Results on constraint-free engineering.} To assess whether models can engineer a protein without expert guidance on where to intervene, we evaluate the constraint-free variant of the engineering queries introduced in \cref{4.2} under the same tool-augmented setting. As shown in \cref{tab:constraint-free-engineering}, removing the defective and protected positions lowers the pass rate of four out of five models, from 45.0--65.0 to 30.0--55.0. The two largest drops arise from opposite failure modes. Specifically, Qwen3.5-397B-A17B applies generic heuristics such as removing Gly/Cys or substituting alanine without structural verification, which reduces in silico validity. DeepSeek-V3.2 instead withholds answers, with empty responses rising from 8/60 to 24/60 and its median tool calls from 5 to 14 as it searches without committing to a candidate. The difficulty of engineering therefore lies not only in choosing how to mutate a residue, but also in identifying which residues to modify.

\noindent\textbf{Results of non-LLM generative models.} To examine whether the difficulty of \system{} extends beyond LLMs, we evaluate three multimodal text--protein generation models, ProteinDT, PAAG, and ProDVa, which map short functional descriptions to sequences through contrastive text--protein alignment or fragment retrieval~(\cref{tab:non-llm-multimodal-rubric-summary}). Despite their domain-specific architectures, all three score zero on engineering, as they treat engineering queries as de novo design requests and emit sequences distinct from the wild-type. In generation, ProteinDT and PAAG produce highly valid and novel sequences, but their outputs rarely fold or function. Two established binder design models, BindCraft and BoltzGen, predict well-folded binder complexes (91.1 and 95.8 fold validity) and surpass all LLM baselines on binder queries (\cref{tab:generation-rubric-breakdown-binder}), yet their pass rates remain at 20.0 and 36.7, bounded by functional consistency. \system{} therefore remains challenging even for models purpose-built for protein design.

\begin{figure}[t!]
    \centering
    \includegraphics[width=\linewidth]
    {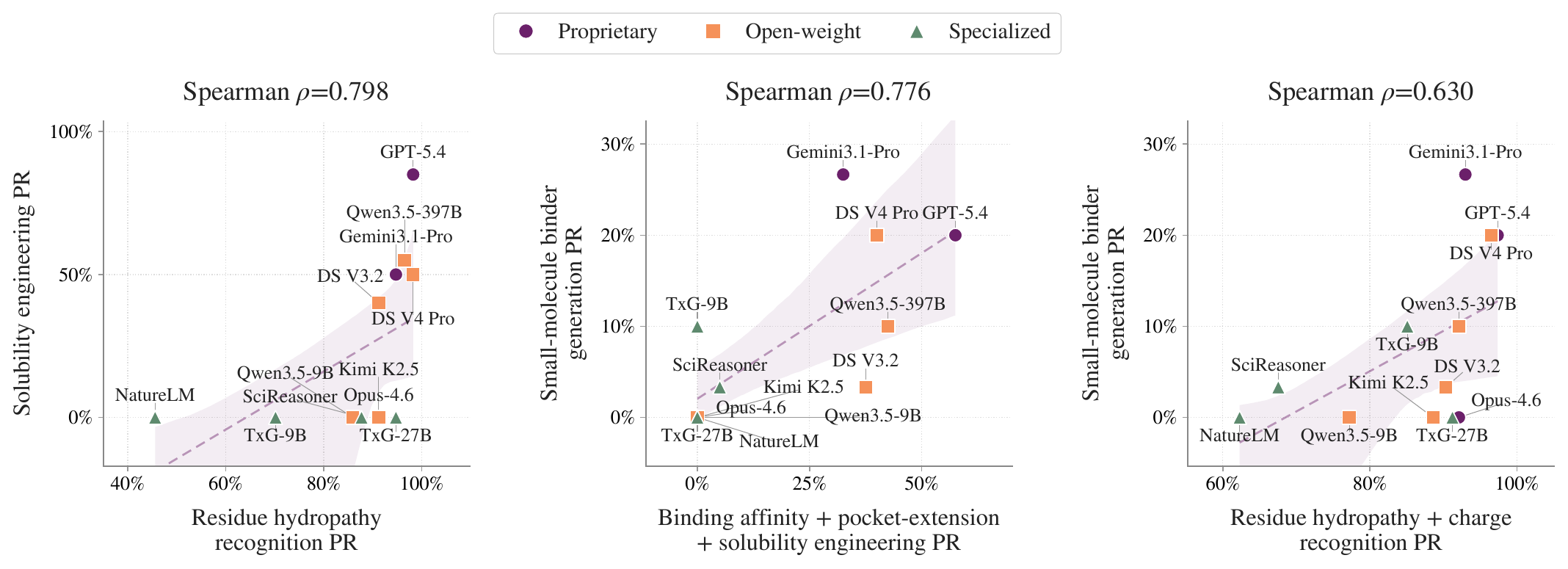}
    \caption{
    Cross-stage correlations between recognition, engineering, and generation performance of baseline LLMs. DS denotes DeepSeek and TxG indicates TxGemma-chat models.
    }
    \label{fig:task-correlation}
    \vskip -10pt
\end{figure}

\noindent\textbf{Cross-stage correlation.}
Our three-stage design raises the question of whether model performance is associated across stages. We therefore compute Spearman correlations between the pass rates (PR) of biologically motivated subtask pairings. We pair residue hydropathy recognition with solubility engineering, since the latter requires identifying residues whose hydropathy creates surface aggregation liabilities. We pair the aggregate of solubility, binding affinity, and pocket-extension engineering with small-molecule binder generation, since these tasks involve complementary aspects of surface hydropathy, chemical complementarity, and pocket geometry. Finally, we pair residue hydropathy and charge recognition with small-molecule binder generation to examine a direct recognition--generation association. As \cref{fig:task-correlation} shows, all three pairs correlate positively, which is consistent with shared physicochemical demands across stages. These cross-stage associations provide empirical support for our benchmark’s three-stage design, which connects foundational recognition skills with engineering and generation in an integrated assessment of protein-design competence. \cref{fig:task-correlation-appendix} reports additional biologically linked pairings.

\begin{wrapfigure}{r}{0.5\linewidth}
    \vspace{-\baselineskip}
    \centering
    \includegraphics[width=\linewidth]{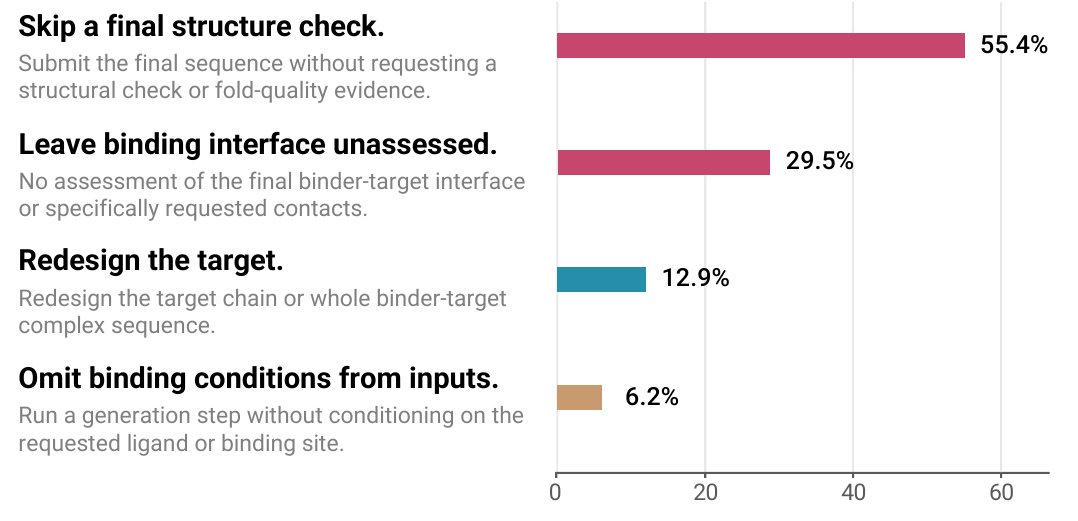}
    \caption{Top four recurring workflow-level failure patterns of the tool-augmented agents. Each trajectory can exhibit multiple patterns, and the reported shares are therefore not mutually exclusive.}
    \label{fig:failure-analysis}
    \vskip -10pt
\end{wrapfigure}

\noindent\textbf{Failure analysis.}
To characterize recurring problems in the tool-augmented agent setting, we use GPT-5.6-Sol~\citep{openai2026gpt56} to annotate each trajectory from the five models in \cref{tab:tool-rubric} with the workflow-level patterns it exhibits. As \cref{fig:failure-analysis} shows, the most frequent pattern is submitting a sequence without any structural check (55.4\%), followed by leaving the binder--target interface unassessed (29.5\%). On miniprotein binder generation for 9DQ0, for instance, DeepSeek-V4-Pro folds the binder alone (pLDDT 86.2) and submits it without assessing its interaction with the target. Beyond validation, agents also redesign the target chain rather than the binder (12.9\%) or drop the requested ligand or binding-site conditioning from a generation step (6.2\%). The latter arises either from unconditional backbone generation, or from conditioning one tool on the requested sites and then submitting a candidate produced by a second generation tool called without that conditioning. Agentic protein design therefore demands both explicit guidance on which properties to verify and a mechanism that maintains the requested constraints attached to the sequence finally submitted, where tool access alone provides neither. Additional examples in \cref{tab:agent-failure-examples}.



\section{Conclusion}\label{06_conclusion}

We presented \system{}, a language-interfaced benchmark that probes generalist protein design capabilities through three complementary stages: recognition, rationale-guided engineering, and functional generation. 
Our evaluation shows that generalist vibe protein design remains a substantial open challenge, motivating language-interfaced models that integrate biological understanding, mechanistic reasoning, and grounded generation within a unified framework.

\subsubsection*{Acknowledgments}
This work was supported by the Ministry of Science and ICT (MSIT), Republic of Korea, through the National IT Industry Promotion Agency (NIPA), as part of the Domain-Specific Foundation Model Project (Grant No. PJT-26-100004). Additionally, we thank Woosung Jeon, Seri Jo, and Jason Yang for valuable feedback and careful review of our benchmark queries.

\bibliography{iclr2027_conference}
\bibliographystyle{iclr2027_conference}

\newpage
\appendix
\section*{Supplementary Materials}

\section{Experimental setting}\label{A}

We evaluate all LLMs in a zero-shot setting using benchmark queries as the user prompt, without task-specific fine-tuning or in-context examples. GPT-5.4~\citep{gpt-5}, Gemini3.1-Pro~\citep{gemini-3.1}, Claude Opus-4.6~\citep{opus4.6}, DeepSeek-V4-Pro~\citep{deepseekv4}, DeepSeek-V3.2~\citep{deepseekv3.2}, Kimi-K2.5~\citep{kimi-k2.5}, and Qwen3.5~\citep{qwen3.5} are accessed via the OpenRouter API. NatureLM-8$\times$7B~\citep{naturelm}, SciReasoner-8B~\citep{scireasoner}, and TxGemma~\citep{txgemma} are run locally through HuggingFace Transformers with SGLang~\citep{sglang}. Unless otherwise specified, decoding uses deterministic inference with temperature 0 and a maximum of 65{,}536 new tokens, with up to five retries using exponential backoff. We enable internal reasoning (\ie, thinking tokens) for GPT-5.4, Gemini3.1-Pro, DeepSeek-V4-Pro, DeepSeek-V3.2, Kimi-K2.5, and Qwen3.5-397B-A17B. For Opus-4.6 and Qwen3.5-9B, however, we disable reasoning, as enabling it yielded an excessive number of empty responses. Local models are run in bfloat16. For non-LLM multimodal baselines, we use the released checkpoints and inference implementations without task-specific fine-tuning. PAAG~\citep{paag} is evaluated through its native text-conditioned sequence generation interface with stochastic sampling. ProDVa~\citep{ProDVa} uses the released inference configuration: stochastic decoding with temperature 0.7, top-k 950, and a maximum of 256 new tokens. ProteinDT-T5~\citep{proteindt} follows the released T5 decoder configuration with temperature 1.0, top-k 40, top-p 0.9, repetition penalty 1.0, and one beam; we generate 8 candidates and select the one with the highest CLAP similarity. For non-LLM binder generation baselines, we use structure-conditioned binder design pipelines for BoltzGen~\citep{boltzgen} and BindCraft~\citep{bindcraft}, with 500 diffusion steps for BoltzGen and up to 140 sequence-optimization iterations across four stages (75, 45, 5, and 15 iterations, respectively) for BindCraft. BindCraft is evaluated on the 90 protein-target binder queries, whereas BoltzGen additionally supports the 30 small-molecule-target queries. We provide the corresponding targets and binding-site constraints, with binder lengths restricted to 40--80 residues for miniproteins and 30--150 residues otherwise. Both baselines request one design per query without benchmark-score-based candidate selection.

For the tool-augmented setting, we use the ReAct~\citep{react} framework with a maximum of 30 iterations per query. Each model has access to the same tool set spanning protein design, validation, analysis, and workflow management:
\begin{itemize}[topsep=0pt,itemsep=1mm, parsep=0pt, leftmargin=5mm]
    \item Backbone generation and redesign: RFdiffusion~\citep{rfdiffusion}, RFdiffusion3~\citep{rfdiffusion3}
    \item Inverse folding and sequence redesign: ProteinMPNN~\citep{proteinmpnn}, LigandMPNN~\citep{ligandmpnn}
    \item Sequence masked-redesign: ESM2~\citep{esm2}
    \item Binder design and redesign: BoltzGen~\citep{boltzgen}, PXDesign~\citep{pxdesign}
    \item Structure prediction: Boltz2~\citep{boltz2}, ESMFold2~\citep{esmfold2}
    \item Small-molecule docking: AutoDock Vina~\citep{autodock}
    \item Physics-based scoring and structural relaxation: PyRosetta~\citep{pyrosetta}
\end{itemize}

We additionally provide analysis tools: \texttt{family\_profile} (sequence composition, conservation, and candidate similarity to a family profile), \texttt{sequence\_analysis} (composition, charge, motifs, residue properties, and sequence liabilities), \texttt{structure\_analysis} (residue exposure, contacts, interfaces, and structural liabilities), \texttt{protein\_qc} (sequence-level biophysical properties and liabilities), \texttt{residue\_lookup} (residue identities and properties at specified positions), and \texttt{tmalign} (pairwise structural alignment and similarity).

Workflow tools include \texttt{sequence\_edit} (applies specified substitutions and checks wild-type residue consistency), \texttt{merge\_candidates} (merges and deduplicates candidate lists), \texttt{objective\_rank} (ranks candidates using agent-specified criteria), and \texttt{search\_sequence} (retrieves sequences matching GO or InterPro terms, with a default cutoff retaining UniProt entries dated on or before 2025-09-01).

\begin{table}[t]
\centering
\setlength{\tabcolsep}{5pt}
\renewcommand{\arraystretch}{1.1}
\caption{Examples of recurring workflow patterns in failed ReAct trajectories.
Each trajectory can exhibit multiple patterns, and the reported shares are therefore not
mutually exclusive.}
\label{tab:agent-failure-examples}
\begin{tabular}{>{\raggedright\arraybackslash}p{0.20\textwidth}
                >{\raggedright\arraybackslash}p{0.56\textwidth}
                r}
\toprule
Patterns & Examples & Share (\%) \\
\midrule
Skip a fold check
& \textbf{GPT-5.4 / 4AT2:}
Used \texttt{residue\_lookup} to inspect the input sequence,
then \texttt{sequence\_edit} to introduce five substitutions.
Submitted the resulting 490-residue mutant directly,
without predicting or checking its structure.
& 55.4 \\
\addlinespace
Leave binding interface unassessed
& \textbf{DeepSeek-V4-Pro / 9DQ0:}
Folded the target with ESMFold2 and used PXDesign to generate
a 60-residue binder against that structure.
Folded the binder alone with ESMFold2 (pLDDT 86.2),
then submitted it without assessing the binder--target interface.
& 29.5 \\
\addlinespace
Redesign the target
& \textbf{Qwen3.5-397B / 7M59:}
Used RFdiffusion3 to generate a 60-residue binder backbone
alongside the 65-residue target.
Passed the entire complex to ProteinMPNN without restricting
redesign to the binder chain, changing both chains,
then submitted the combined 125-residue output as the binder.
& 12.9 \\
\addlinespace
Omit binding conditions from design inputs
& \textbf{Qwen3.5-397B / 8HE6:}
Generated a backbone with RFdiffusion3 conditioned on ten
target-site residues, then a sequence with ProteinMPNN.
Subsequently started a separate PXDesign run from the original
target structure without specifying those site constraints,
and submitted its new 60-residue candidate instead.
& 6.2 \\
\addlinespace
Pass JSON/TSV files as FASTA inputs
& \textbf{DeepSeek-V4-Pro / 2ZEJ:}
Retrieved candidates with \texttt{search\_sequence}, then passed
the returned JSON file to \texttt{family\_profile} through
\texttt{alignment\_path}, which explicitly requires aligned FASTA.
After the file-format rejection, submitted its final answer
without retrying the tool.
& 4.3 \\
\addlinespace
Pass structure or search-result files as YAML configurations.
& \textbf{Qwen3.5-397B / CID 177867006:}
Obtained a structure from Boltz2, then passed its mmCIF file
to boltzgen through \texttt{target\_yaml},
which explicitly requires a YAML design specification.
After the parsing error, switched to PXDesign and RFdiffusion3
without correcting the BoltzGen input.
& 1.5 \\
\bottomrule
\end{tabular}
\vskip -10pt
\end{table}
\section{Additional Experimental Results}\label{B}

\paragraph{Trajectory examples for each workflow failure pattern.}
We present per-pattern examples of the recurring workflow patterns of
\cref{fig:failure-analysis} in \cref{tab:agent-failure-examples}. GPT-5.4 introduces five substitutions and submits the resulting mutant without ever predicting its structure, and DeepSeek-V4-Pro obtains a confidently folded binder yet never evaluates the complex it was asked to bind. In the remaining two examples, the design object itself drifts from the specification, as ProteinMPNN is applied to the entire complex rather than the binder chain, and a second generation run is launched without the site constraints that conditioned the first. Additionally, agents pass JSON/TSV files as FASTA inputs (4.3\%) or structure/search-result files as YAML configurations (1.5\%), but these two argument-specification errors are less frequent than the aforementioned four patterns. Overall, these examples motivate explicit guidance on which properties to verify and a mechanism that keeps the requested constraints attached to the sequence finally submitted.

\begin{table}[t]
\centering
\small
\caption{
Generation performance and token usage of ReAct and advanced Codex CLI harness on the same 30-query subset.
Performance values are percentages; token counts are summed over all 30 queries and reported in millions.
}
\label{tab:codex-harness}
\setlength{\tabcolsep}{4pt}
\resizebox{\textwidth}{!}{
\begin{tabular}{l l c c c >{\columncolor{piggypink!50}}c r r r >{\columncolor{piggypink!50}}r}
\toprule
& & \multicolumn{4}{c}{Generation}
& \multicolumn{4}{c}{Token usage (M)} \\
\cmidrule(lr){3-6}
\cmidrule(lr){7-10}
Models & Setup
& Seq. valid. & Fold valid. & Func. Cons. & Pass rate
& Input & Output & Total & vs. ReAct \\
\midrule
\multirow{2}{*}{DeepSeek-V4-Pro}
& ReAct (30 iterations) & 100.0 & 93.3 & 46.7 & 43.3 & \pz\pz5.3 & \pz0.3 & \pz\pz\pz5.6 & -- \\
& Codex (iteration-free) & 100.0 & 90.0 & 56.7 & \textbf{50.0} & 904.4 & \pz7.3 & \pz911.8 & \textbf{$\times$163.6} \\
\arrayrulecolor{black!40}\midrule
\arrayrulecolor{black}
\multirow{2}{*}{DeepSeek-V3.2}
& ReAct (30 iterations) & \pz63.3 & 50.0 & 26.7 & 26.7 & \pz\pz9.5 & \pz0.2 & \pz\pz\pz9.7 & -- \\
& Codex (iteration-free) & 100.0 & 90.0 & 33.3 & \textbf{30.0} & 1{,}309.7 & 12.6 & 1{,}322.2 & \textbf{$\times$136.6} \\
\arrayrulecolor{black!40}\midrule
\arrayrulecolor{black}
\multirow{2}{*}{Qwen3.5-397B-A17B}
& ReAct (30 iterations) & \pz73.3 & 53.3 & 13.3 & 13.3 & \pz\pz3.0 & \pz0.1 & \pz\pz\pz3.1 & -- \\
& Codex (iteration-free) & 100.0 & 90.0 & 43.3 & \textbf{40.0} & 308.4 & \pz1.2 & \pz309.6 & \textbf{$\times$101.0} \\
\bottomrule
\end{tabular}
}
\vskip -5pt
\end{table}
\paragraph{Effect of unbounded iteration with an advanced agent setup.} To test whether the remaining failures in \cref{tab:tool-rubric} reflect a limited iteration budget and agent setup rather than a limit of design competence, we re-run the 30-query generation subset under an advanced Codex CLI harness~\citep{openai_codex_cli}, which exposes the same tool list as ReAct but adds skills instructing the agent to keep iterating until its acceptance criteria are met instead of terminating after a failed attempt. As shown in \cref{tab:codex-harness}, this raises the pass rate for all three models, from 43.3 to 50.0, from 26.7 to 30.0, and from 13.3 to 40.0, but sacrifices at 101.0 to 163.6 times the token cost. Despite this increase in token usage, the best pass rate remains 50.0 for these three models, indicating that neither an advanced agent setup nor a greatly enlarged iteration budget is sufficient to solve vibe protein design.

\begin{table}[t]
\centering
\small
\caption{
Selected observable workflow failures and top-five tools of ReAct-based DeepSeek-V4-Pro and Qwen3.5-397B-A17B on generation tasks. This comparison highlights primary failure categories with higher shares in Qwen3.5-397B-A17B.
}
\label{tab:agent-errors}
\setlength{\tabcolsep}{6pt}
\renewcommand{\arraystretch}{1.15}
\begin{tabular}{@{}l c c@{}}
\toprule
\multicolumn{3}{@{}l}{%
\emph{Selected observable workflow failures (\%)}} \\
\midrule
Failure types & DeepSeek-V4-Pro & Qwen3.5-397B-A17B \\
\midrule
End without an evaluable final sequence
& 5.2 & \textbf{36.7} \\
Submit the final sequence without directly validating it
& 20.0 & \textbf{48.1} \\
Inverse-fold both target and binder to produce the final answer
& 0.0 & \textbf{31.1} \\
Finish without obtaining the requested site-contact evidence
& 10.0 & \textbf{93.3} \\
\midrule
\multicolumn{3}{@{}l}{%
\emph{Top five tools by query-level usage}} \\
\midrule
\multicolumn{3}{@{}>{\raggedright\arraybackslash}p{\linewidth}@{}}{%
\textbf{DeepSeek-V4-Pro:}
\texttt{protein\_qc}, \texttt{esmfold2},
\texttt{sequence\_analysis}, \texttt{search\_sequence},
\texttt{pxdesign}} \\
\addlinespace
\multicolumn{3}{@{}>{\raggedright\arraybackslash}p{\linewidth}@{}}{%
\textbf{Qwen3.5-397B-A17B:}
\texttt{esmfold2}, \texttt{search\_sequence},
\texttt{inverse\_folding}, \texttt{rfdiffusion3},
\texttt{pxdesign}} \\
\bottomrule
\end{tabular}
\vskip -5pt
\end{table}
\begin{figure}[t]
\centering
\begin{tcbraster}[
  raster columns=2, raster equal height=rows,
  raster column skip=3mm,
  raster before skip=0pt, raster after skip=0pt,
  raster left skip=0pt, raster right skip=0pt]
\begin{tcolorbox}[
  colback=vpteal!6, colframe=vpteal!65!black,
  boxrule=0.4pt, arc=1pt,
  left=3pt, right=3pt, top=3pt, bottom=3pt,
  title={DeepSeek-V4-Pro / 7M59 \textbullet\ Pass},
  fonttitle=\footnotesize]
\scriptsize
\textbf{Task:} Design a binder contacting at least
half of the specified target-site residues: I32, Q5, Y38, R12,
T31, D7, V3, M20, Q16, and I4.
\par\smallskip
\ttfamily
[2] rfdiffusion(\\
\hspace*{1em}target\_chain="A",\\
\hspace*{1em}\textcolor{vpteal!80!black}{target\_hotspots=[32,5,38,\ldots,4]}, \ldots)\\
$\rightarrow$ hotspot-conditioned complex backbone\\[3pt]
[3] inverse\_folding(\\
\hspace*{1em}\textcolor{vpteal!80!black}{chains=["B"]},\\
\hspace*{1em}pdb\_path=\ldots, num\_designs=16, \ldots)\\
$\rightarrow$ \textcolor{vpteal!80!black}{65-aa binder sequence: MVV\ldots GTR}\\[3pt]
[9] esmfold2(\\
\hspace*{1em}target\_sequence=\ldots,\\
\hspace*{1em}\textcolor{vpteal!80!black}{binder\_sequence=MVV\ldots GTR}, \ldots)\\
$\rightarrow$ pLDDT=80.42, pTM=0.755, iPTM=0.704\\[3pt]
[10] structure\_analysis(\\
\hspace*{1em}interface\_chain\_ids=["A","B"],\\
\hspace*{1em}\textcolor{vpteal!80!black}{positions=[3,4,5,\ldots,38]}, \ldots)\\
$\rightarrow$ site\_contacts.requested\_position\_count=10\\
$\rightarrow$ \textcolor{vpteal!80!black}{site\_contacts.contact\_count=9}\\[3pt]
{\normalfont
\textbf{Reason:} \ldots used hotspot-guided binder design \ldots
inverse-folded only the binder chain \ldots
the submitted sequence directly contacts 9 of the 10 site residues \ldots\\
\textbf{Answer:} MVVVELVL\ldots APPYDGTR
}
\end{tcolorbox}
\begin{tcolorbox}[
  colback=vprose!5, colframe=vprose!80!black,
  boxrule=0.4pt, arc=1pt,
  left=3pt, right=3pt, top=3pt, bottom=3pt,
  title={Qwen3.5-397B-A17B / 7M59 \textbullet\ Fail},
  fonttitle=\footnotesize]
\scriptsize
\textbf{Task:} Design a binder contacting at least
half of the specified target-site residues: I32, Q5, Y38, R12,
T31, D7, V3, M20, Q16, and I4.
\par\smallskip
\ttfamily
[2] rfdiffusion3(binder\_length=60,\\
\hspace*{1em}\textcolor{vpteal!80!black}{target\_hotspots=[32,5,38,\ldots,4]},\\
\hspace*{1em}target\_pdb=\ldots, \ldots)\\
$\rightarrow$ complex backbone: A=60 aa binder, B=65 aa target\\[3pt]
[3] inverse\_folding(\textcolor{vprose!80!black}{chains=None},\\
pdb\_path=\ldots,\hspace*{1em}num\_designs=8)\\
{\normalfont\itshape\color{vprose!80!black}
[No binder-chain restriction is supplied.]}\\
$\rightarrow$ \textcolor{vprose!80!black}{125-aa whole-complex sequence}\\[3pt]
{\normalfont
\textbf{Reason:} \ldots designed sequences for the binder backbone \ldots
The top-ranked sequence was selected.\\
\textbf{Answer:} AATPEEIA\ldots LPDYPGDR\\
{\normalfont\itshape\color{vprose!80!black}
[Answer is the entire A+B output from call 3.]}
}
\end{tcolorbox}
\end{tcbraster}
\vspace{7pt}
\caption{
Matched trajectories for the same binding-site-conditioned generation query.
}
\label{fig:agent-trajectories}
\vskip -10pt
\end{figure}
\paragraph{What separates a stronger agent from a weaker one.} To identify what distinguishes agents that succeed under tool use from those that do not, we use GPT-5.6-Sol~\citep{openai2026gpt56} to characterize representative failures in the ReAct trajectories of DeepSeek-V4-Pro (generation pass rate 23.8) and Qwen3.5-397B-A17B (6.2). \cref{tab:agent-errors} shows that the gap is one of verification rather than tool choice. The two agents share three of their five most-used tools, and the tools that distinguish them divide along functional lines: DeepSeek-V4-Pro additionally relies on \texttt{protein\_qc} and
\texttt{sequence\_analysis}, which inspect intermediate outputs, whereas Qwen3.5-397B-A17B relies on \texttt{inverse\_folding} and \texttt{rfdiffusion3}, which produce them. Moreover, Qwen3.5-397B-A17B finishes without obtaining the requested site-contact evidence in 93.3\% of its generation queries against 10.0\% for DeepSeek-V4-Pro, and submits a final sequence
without directly validating it in 48.1\% against 20.0\%. It also loses the design object outright, inverse-folding target and binder together to form the answer in 31.1\% of queries where DeepSeek-V4-Pro never does so, and ends without an evaluable sequence in 36.7\% against 5.2\%.

The examples in \cref{fig:agent-trajectories} illustrate these differences. On 7M59, DeepSeek-V4-Pro restricts inverse folding to the 65-residue binder, verifies contacts with 9 of 10 requested residues, validates the final sequence, and submits a binder that passes the benchmark. Qwen instead inverse-folds the 60-residue binder and 65-residue target together, then submits the complete 125-residue output as the binder. Although all of its tool calls succeed, it neither checks the requested contacts nor validates the submitted sequence. This suggests that what sets a stronger agent apart is not its choice of tools but its ability to preserve the design object and verify task-specific requirements before submission.

\begin{table}[t]
\centering
\small
\setlength{\tabcolsep}{5pt}
\renewcommand{\arraystretch}{1.15}
\caption{Non-default tool arguments requested by tool-augmented agents. Share is the percentage of all tool-augmented trajectories in which the pattern occurs.}
\label{tab:tool-arguments}
\begin{tabular}{l >{\raggedright\arraybackslash}p{0.60\textwidth} r}
\toprule
Arguments & \normalsize Examples & Share \\
\midrule
Fewer diffusion steps & 400 / 200 / 500 $\rightarrow$ 20--100 (PXDesign, RFdiffusion3, BoltzGen) & 24.1 \\
Lower retrieval cap & \texttt{max\_sequences} 100 $\rightarrow$ 1--10;\newline \texttt{max\_downstream\_candidates} 10 $\rightarrow$ 1 & 12.5 \\
More designs per call & \texttt{num\_samples} 5 $\rightarrow$ 16--512 (PXDesign) & 10.3 \\
Fewer designs per call & 8 / 10 / 8 / 10 $\rightarrow$ 1--2 (MPNN, RFdiffusion, RFdiffusion3, BoltzGen) & 10.2 \\
Higher sampling temperature & 0.1 $\rightarrow$ 0.5--1.0 (MPNN); 1.0 $\rightarrow$ 1.2 (PXDesign diffusion) & \pz8.8 \\
Higher retrieval cap & \texttt{max\_sequences} 100 $\rightarrow$ 500--2{,}000;\newline \texttt{max\_downstream\_candidates} 10 $\rightarrow$ 50--500 & \pz6.0 \\
\bottomrule
\end{tabular}
\vskip -10pt
\end{table}
\paragraph{Tool argument selection.} We characterize how agents configure the tools they call by categorizing, with GPT-5.6 Sol, every request that departs from a tool's default. As shown in \cref{tab:tool-arguments}, agents most often reduce the parameters that govern a tool's compute budget, suggesting that they optimize for speed rather than design quality. Cutting the number of diffusion steps is the most frequent override, appearing in 24.1\% of trajectories: agents request 20--100 steps where PXDesign, RFdiffusion3, and BoltzGen default to 400, 200, and 500. At the same time, they pursue sequence diversity through the settings, raising the sampling temperature in 8.8\% of trajectories. Search breadth shows a similar bias toward narrowing, as the cap on retrieved candidates is lowered about twice as often as it is raised (12.5\% versus 6.0\%). Overall, these tendencies indicate that agents consistently trade design quality for speed and diversity.

\definecolor{vpgreen}{HTML}{1F5E4E}   
\definecolor{vpkey}{HTML}{0F6E8C}     

\begin{figure}[t]
\centering
\begin{tcbraster}[
  raster columns=1,
  raster row skip=3mm,
  raster before skip=0pt, raster after skip=0pt,
  raster left skip=0pt, raster right skip=0pt]
\begin{tcolorbox}[
  colback=vpgreen!6, colframe=vpgreen!70!black,
  boxrule=0.4pt, arc=1pt,
  left=3pt, right=3pt, top=3pt, bottom=3pt,
  title={DeepSeek-V4-Pro / GO:0005267 \textbullet\ Tool-free},
  fonttitle=\footnotesize]
\scriptsize
\textbf{Task:} Design a protein sequence satisfying the molecular function
\emph{potassium channel activity}: enables the energy-independent facilitated diffusion
of a potassium ion through a transmembrane aqueous pore or channel.
\par\smallskip
\textbf{Reason:} \ldots the characteristic K+ channel fold with
\textcolor{vpkey}{two transmembrane helices (M1 and M2) flanking a re-entrant pore loop}
\ldots The pore loop contains a short \textcolor{vpkey}{pore helix (ALWSTVEA)} and the
highly conserved \textcolor{vpkey}{selectivity filter sequence TVGYG}, whose
\textcolor{vpkey}{backbone carbonyls coordinate dehydrated K+ ions}, enabling
energy-independent diffusion. The sequence is designed to
\textcolor{vpkey}{tetramerize}, forming a central ion conduction pore.
\par\smallskip
\textbf{Answer:} MLVGLV\ldots LWSTEA\textcolor{vpkey}{TVGYG}DLYPVT\ldots VGREQ
\end{tcolorbox}
\begin{tcolorbox}[
  colback=vpteal!6, colframe=vpteal!65!black,
  boxrule=0.4pt, arc=1pt,
  left=3pt, right=3pt, top=3pt, bottom=3pt,
  title={DeepSeek-V4-Pro / GO:0005267 \textbullet\ ReAct},
  fonttitle=\footnotesize]
\scriptsize
\ttfamily
{\normalfont
\textbf{Hidden reasoning:} A minimal K+ channel could be designed based on the
\textcolor{vpkey}{KcsA-like architecture: two transmembrane helices (TM1, TM2), a pore
helix (P), and the selectivity filter with the signature GYG sequence}. \ldots I could
use search\_sequence to retrieve real potassium channel sequences and use
family\_profile to build a profile, then use that to guide a de novo design.
}\\[3pt]
[1] search\_sequence(\textcolor{vpkey}{go\_term="GO:0005267"}, reviewed=true,
max\_downstream\_candidates=40)\\
$\rightarrow$ 21 candidates with full sequences\\[3pt]
[2] family\_profile(candidates=\ldots, align=true, min\_conservation=0.75,
min\_coverage=0.5, \ldots)\\
$\rightarrow$ MAFFT alignment of 21 reference sequences\\
$\rightarrow$ consensus motif: \textcolor{vpkey}{TIGYGD}\\
$\rightarrow$ conserved positions ($\geq75\%$; alignment columns):
\textcolor{vpkey}{T(864), G(866), G(868), D(869)}\\[3pt]
{\normalfont
\textbf{Hidden reasoning:} \ldots The conserved positions listed: T(864), G(866),
G(868), D(869). The classic selectivity filter is \textcolor{vpkey}{TVGYG}. \ldots KcsA
sequence (for reference, I know it roughly): \ldots I shouldn't just copy KcsA (that's
not de novo). However, the selectivity filter \textcolor{vpkey}{TVGYG must be present}
\ldots
}\\[3pt]
{\normalfont
\textbf{Reason:} Potassium channels share a minimal architecture: an
\textcolor{vpkey}{outer transmembrane helix (M1)}, a turret, a
\textcolor{vpkey}{short pore helix}, the signature
\textcolor{vpkey}{TVGYG selectivity filter} \ldots and an
\textcolor{vpkey}{inner transmembrane helix (M2)}. I authored a de novo sequence that
assembles this \textcolor{vpkey}{KcsA-like topology}, embedding the conserved
\textcolor{vpkey}{TVGYG selectivity filter} and placing hydrophobic membrane-spanning
helices around it.\\
\textbf{Answer:} MALLGL\ldots\textcolor{vpkey}{TVGYG}DLYPVT\ldots WFVGR
}
\end{tcolorbox}
\end{tcbraster}
\vspace{7pt}
\caption{
The same model on the same GO-conditioned generation query under tool-free (top) and
ReAct-based tool-augmented (bottom) settings. Highlighted text marks the mechanistic
commitment that recurs across both settings, the TVGYG selectivity filter and the
two-transmembrane-helix pore architecture.
}
\label{fig:cross-setting}
\vskip -15pt
\end{figure}

\noindent\textbf{Continuity between the tool-free and tool-augmented settings.}
Here, we show that evaluating LLMs both with and without tools as done in our evaluations is necessary, as a model's internal design knowledge shapes its behavior once tools are available. To this end, we compare the tool-free and ReAct trajectories of one model on a single GO-conditioned generation query. Without tools, on potassium channel activity (GO:0005267), DeepSeek-V4-Pro names the two-transmembrane-helix pore architecture and the TVGYG selectivity filter, and attributes energy-independent conduction to backbone carbonyls coordinating dehydrated K$^{+}$ ions. Given tools on the identical prompt, the same commitments determine its actions: it retrieves reviewed sequences for the corresponding GO term, builds an alignment-based family profile, and combines the resulting conservation evidence with its prior knowledge of the TVGYG filter and KcsA-like architecture to construct a candidate sequence (\cref{fig:cross-setting}). Both the translation of the semantic objective into a tool query and the interpretation of the returned evidence draw on what the model already knows about the target function. This case shows that competence observed without tools strongly shapes agent behavior once tools are available. Evaluating both settings therefore allows us to assess intrinsic protein reasoning alongside its translation into agentic protein design.

\noindent\textbf{Results for antibody, DNA, and RNA binder generation.}
\begin{table}[t]
\centering
\caption{
Additional results on antibody, RNA, and DNA binder generation in no-tool setting.
}
\label{tab:additional-binder-generation}
\resizebox{\textwidth}{!}{
\begin{tabular}{l c c c c c c c c c c}
\toprule
& \multicolumn{3}{c}{Antibody CDR-H3}
& \multicolumn{3}{c}{RNA binder}
& \multicolumn{3}{c}{DNA binder}
& \\
\cmidrule(lr){2-4}
\cmidrule(lr){5-7}
\cmidrule(lr){8-10}
Models
& \makecell{Seq.\\valid.}
& \makecell{Fold\\valid.}
& \makecell{Func.\\Cons.}
& \makecell{Seq.\\valid.}
& \makecell{Fold\\valid.}
& \makecell{Func.\\Cons.}
& \makecell{Seq.\\valid.}
& \makecell{Fold\\valid.}
& \makecell{Func.\\Cons.}
& Overall \\
\midrule
\multicolumn{11}{>{\columncolor{piggypink!50}}l}
{\emph{General-purpose LLMs}\vspace{0.02in}} \\
GPT-5.4
& 100.0 & 96.7 & 3.3
& 96.7 & 3.3 & 0.0
& 80.0 & 73.3 & 46.7
& \underline{16.7} \\
Gemini3.1-Pro
& 100.0 & 100.0 & 0.0
& 96.7 & 60.0 & 3.3
& 96.7 & 96.7 & 66.7
& \textbf{23.3} \\
Opus-4.6
& 46.7 & 43.3 & 0.0
& 73.3 & 16.7 & 0.0
& 100.0 & 70.0 & 33.3
& 11.1 \\
DeepSeek-V4-Pro
& 100.0 & 96.7 & 0.0
& 100.0 & 36.7 & 10.0
& 100.0 & 70.0 & 33.3
& 14.4 \\
DeepSeek-V3.2
& 100.0 & 100.0 & 10.0
& 100.0 & 50.0 & 3.3
& 100.0 & 33.3 & 6.7
& 6.7 \\
Kimi-K2.5
& 100.0 & 96.7 & 0.0
& 83.3 & 20.0 & 0.0
& 93.3 & 70.0 & 23.3
& 7.8 \\
Qwen3.5-397B-A17B
& 100.0 & 96.7 & 0.0
& 96.7 & 50.0 & 0.0
& 100.0 & 76.7 & 10.0
& 3.3 \\
Qwen3.5-9B
& 60.0 & 56.7 & 3.3
& 73.3 & 56.7 & 0.0
& 60.0 & 56.7 & 0.0
& 1.1 \\
\midrule
\multicolumn{11}{>{\columncolor{piggypink!50}}l}
{\emph{Domain-specialized LLMs}\vspace{0.02in}} \\
NatureLM-8$\times$7B
& 100.0 & 96.7 & 0.0
& 70.0 & 33.3 & 0.0
& 73.3 & 20.0 & 0.0
& 0.0 \\
SciReasoner-8B
& 100.0 & 93.3 & 0.0
& 100.0 & 33.3 & 0.0
& 100.0 & 43.3 & 3.3
& 1.1 \\
TxGemma-9B
& 100.0 & 96.7 & 3.3
& 100.0 & 10.0 & 0.0
& 100.0 & 46.7 & 0.0
& 1.1 \\
TxGemma-27B
& 100.0 & 96.7 & 0.0
& 100.0 & 33.3 & 0.0
& 100.0 & 33.3 & 0.0
& 0.0 \\
\arrayrulecolor{black}
\bottomrule
\end{tabular}
}
\vskip -5pt
\end{table}
To cover binder design against a wider range of target types, we extend \system{} with antibody, RNA, and DNA binder queries. These classes could not be included under our original temporal cutoff criterion, as both a SAbDab-based search and an independent RCSB--UniProt search identified no eligible antigen whose UniProt accession was first released on or after September 1, 2025, while an exhaustive search over structurally eligible nucleic-acid targets identified no corresponding RNA or DNA target independently registered after the cutoff in RNAcentral~\citep{rnacentral} or GenBank~\citep{genbank}. 

We therefore report an auxiliary set comprising 30 antibody, 30 RNA-binder, and 30 DNA-binder queries under a structure-based temporal guard. Each query is derived from an X-ray diffraction or electron-microscopy complex first released by RCSB on or after September 1, 2025, with resolution $\leq 3$~\AA{}, and its prompt-visible target sequence or binding-region strand has no gapless, fully identical match in a pre-cutoff RCSB polymer entity. Antibody queries request a CDR-H3 sequence given the fixed heavy- and light-chain context and the antigen, whereas nucleic-acid queries provide the experimentally contacted single- or double-stranded target regions in the 5$'$-to-3$'$ direction.

In this experiment, sequence validity requires a non-empty sequence composed exclusively of the 20 canonical amino acids. For RNA- and DNA-binder queries, outputs composed solely of the corresponding nucleotide alphabet, $\{A,C,G,U\}$ or $\{A,C,G,T\}$, are rejected; no sequence-length threshold is imposed. Fold validity requires the global complex pLDDT to exceed 70, without an additional pTM criterion. Functional consistency is evaluated using Protenix-v1. For RNA and DNA targets, it requires the maximum generated-protein--target-strand chain-pair iPTM to exceed 0.8. For antibody targets, we compute a regional iPTM from the Protenix-v1 PAE-bin probabilities after restricting the evaluated tokens to the designed CDR-H3 and the antigen, and require this score to exceed 0.8. This regional criterion prevents confidence attributable primarily to the fixed antibody framework from satisfying the functional-consistency criterion. A query passes only when sequence validity, fold validity, and functional consistency are all satisfied.

As shown in \cref{tab:additional-binder-generation}, these targets remain challenging for current LLMs, where the best overall pass rate is 23.3 from Gemini3.1-Pro. As in \cref{tab:generation-rubric-breakdown-binder}, functional consistency is the primary factor limiting performance. On antibody CDR-H3, ten of the twelve models attain 93.3--100.0 fold validity yet fall to 0.0--10.0 functional consistency, and on RNA binder generation eight models attain 33.3--60.0 fold validity while none exceeds 10.0. Domain-specialized models fail almost entirely, with
NatureLM-8$\times$7B and TxGemma-27B scoring 0.0 functional consistency on all three targets despite sequence validity of 70.0--100.0. These results indicate that the functional-grounding bottleneck identified on protein binder generation persists regardless of binding targets, and that the protein data in the pretraining mixture of these domain-specialized LLMs does not carry over to interface-level quality.

\begin{table}[t]
\centering
\caption{
Cutoff ablation for binder generation in the no-tool setting, comparing binding targets first registered in UniProt (proteins) or PubChem (small molecules) before September 1, 2021 (pre-cutoff) and after September 1, 2025 (post-cutoff).
}
\label{tab:cutoff-ablation}
\resizebox{\textwidth}{!}{
\begin{tabular}{l c c c >{\columncolor{piggypink!50}}c c c c >{\columncolor{piggypink!50}}c}
\toprule
& \multicolumn{4}{c}{Pre-cutoff}
& \multicolumn{4}{c}{Post-cutoff} \\
\cmidrule(lr){2-5}
\cmidrule(lr){6-9}
Models
& Seq. valid.
& Fold valid.
& Func. Cons.
& Pass rate
& Seq. valid.
& Fold valid.
& Func. Cons.
& Pass rate \\
\midrule
DeepSeek-V4-Pro
& 100.0 & 84.2 & 17.5 & 17.5
& 99.2 & 86.7 & 6.7 & 6.7 \\
DeepSeek-V3.2
& 100.0 & 81.7 & 7.5 & 7.5
& 99.2 & 90.0 & 1.7 & 1.7 \\
Qwen3.5-397B-A17B
& 100.0 & 83.3 & 2.5 & 2.5
& 99.2 & 82.5 & 2.5 & 2.5 \\
\bottomrule
\end{tabular}
}
\vskip -10pt
\end{table}
\paragraph{Effect of the temporal cutoff.} To test whether prior exposure to a target inflates design performance, we compare binder generation on targets first registered in UniProt or PubChem before September 1, 2021 (pre-cutoff) against the post-cutoff targets used in \system{}. As shown in \cref{tab:cutoff-ablation}, all three models attain higher or equal pass rates on pre-cutoff targets. The gap is most pronounced in functional consistency, which falls from 17.5 to 6.7 and 7.5 to 1.7 for DeepSeek-V4-Pro and DeepSeek-V3.2, respectively. This may suggest that prior exposure to a target could confer some advantage for binder design, though the effect is model-dependent.

\begin{table}[t]
\centering
\caption{Pass rates for each sequence-level recognition subtask in tool-free setting.}
\label{tab:recognition-rubric-breakdown-seq}
\resizebox{\textwidth}{!}{
\begin{tabular}{l c c c c c c c}
\toprule
Models & \makecell{Motif\\detection} & \makecell{Charge state\\classification} & \makecell{Hydropathy\\classification} & \makecell{Aromaticity\\prediction} & \makecell{Residue charge\\identification} & \makecell{Residue hydropathy\\identification} & Overall \\
\midrule
\multicolumn{8}{>{\columncolor{piggypink!50}}l}{\emph{General-purpose LLMs}\vspace{0.02in}} \\
GPT-5.4 & 77.0 & 89.5 & 94.7 & 100.0 & 96.5 & 98.2 & \textbf{91.9} \\
Gemini3.1-Pro & 56.8 & 28.1 & 5.3 & 28.1 & 91.2 & 94.7 & 51.0 \\
Opus-4.6 & 75.7 & 70.2 & 89.5 & 94.7 & 93.0 & 91.2 & {85.2} \\
DeepSeek-V4-Pro & 90.5 & 80.7 & 77.2 & 86.0 & 94.7 & 98.2 & \underline{88.0} \\
DeepSeek-V3.2 & 78.4 & 70.2 & 84.2 & 93.0 & 89.5 & 91.2 & 84.1 \\
Kimi-K2.5 & 52.7 & 33.3 & 47.4 & 71.9 & 86.0 & 91.2 & 63.2 \\
Qwen3.5-397B-A17B & 85.1 & 77.2 & 80.7 & 94.7 & 87.7 & 96.5 & 86.9 \\
Qwen3.5-9B & 41.9 & 17.5 & 29.8 & 35.1 & 68.4 & 86.0 & 46.2 \\
\midrule
\multicolumn{8}{>{\columncolor{piggypink!50}}l}{\emph{Specialized LLMs}\vspace{0.02in}} \\
NatureLM-8$\times$7B & 0.0 & 56.1 & 14.0 & 0.0 & 78.9 & 45.6 & 30.9 \\
SciReasoner-8B & 27.0 & 21.1 & 50.9 & 8.8 & 47.4 & 87.7 & 39.8 \\
TxGemma-9B & 21.6 & 61.4 & 36.8 & 47.4 & 100.0 & 70.2 & 54.6 \\
TxGemma-27B & 16.2 & 31.6 & 7.0 & 12.3 & 87.7 & 94.7 & 40.4 \\
\arrayrulecolor{black}\bottomrule
\end{tabular}
}
\vskip -5pt
\end{table}

\begin{table}[t]
\centering
\caption{Pass rates for each structure-level recognition subtask in tool-free setting.}
\label{tab:recognition-rubric-breakdown-struct}
\resizebox{0.8\textwidth}{!}{
\begin{tabular}{l c c c c}
\toprule
Models & \makecell{Secondary structure\\identification} & \makecell{Burial\\classification} & \makecell{Disulfide bond\\identification} & Overall \\
\midrule
\multicolumn{5}{>{\columncolor{piggypink!50}}l}{\emph{General-purpose LLMs}\vspace{0.02in}} \\
GPT-5.4 & 44.0 & 82.4 & 40.0 & \textbf{67.9} \\
Gemini3.1-Pro & 32.0 & 80.4 & 40.0 & 63.0 \\
Opus-4.6 & 44.0 & 80.4 & 60.0 & \textbf{67.9} \\
DeepSeek-V4-Pro & 20.0 & 66.7 & 40.0 & 50.6 \\
DeepSeek-V3.2 & 28.0 & 80.4 & 60.0 & 63.0 \\
Kimi-K2.5 & 28.0 & 80.4 & 20.0 & 60.5 \\
Qwen3.5-397B-A17B & 32.0 & 80.4 & 60.0 & \underline{64.2} \\
Qwen3.5-9B & 12.0 & 60.8 & 40.0 & 44.4 \\
\midrule
\multicolumn{5}{>{\columncolor{piggypink!50}}l}{\emph{Specialized LLMs}\vspace{0.02in}} \\
NatureLM-8$\times$7B & 12.0 & 17.6 & 80.0 & 19.8 \\
SciReasoner-8B & 48.0 & 37.3 & 40.0 & 40.7 \\
TxGemma-9B & 36.0 & 41.2 & 60.0 & 40.7 \\
TxGemma-27B & 8.0 & 23.5 & 60.0 & 21.0 \\
\arrayrulecolor{black}\bottomrule
\end{tabular}
}
\vskip -10pt
\end{table}

\begin{table}[t]
\centering
\caption{Pass rates for each domain/function recognition subtask across general-purpose and domain-specialized LLMs in tool-free setting.}
\label{tab:recognition-rubric-breakdown-func}
\resizebox{\textwidth}{!}{
\begin{tabular}{l c c c c c c c}
\toprule
Models & \makecell{Family\\classification} & \makecell{Superfamily\\classification} & \makecell{GO term\\(MF)} & \makecell{GO term\\(BP)} & \makecell{GO term\\(CC)} & \makecell{Fold\\recognition} & Overall \\
\midrule
\multicolumn{8}{>{\columncolor{piggypink!50}}l}{\emph{General-purpose LLMs}\vspace{0.02in}} \\
GPT-5.4 & 77.3 & 76.6 & 72.7 & 71.4 & 66.7 & 50.0 & \textbf{71.1} \\
Gemini3.1-Pro & 71.2 & 79.7 & 36.4 & 42.9 & 66.7 & 27.8 & \underline{62.6} \\
Opus-4.6 & 57.6 & 73.4 & 54.5 & 42.9 & 33.3 & 47.2 & {59.9} \\
DeepSeek-V4-Pro & 63.6 & 68.8 & 36.4 & 42.9 & 0.0 & 36.1 & 56.7 \\
DeepSeek-V3.2 & 54.5 & 53.1 & 36.4 & 42.9 & 33.3 & 33.3 & 48.1 \\
Kimi-K2.5 & 56.1 & 59.4 & 18.2 & 42.9 & 0.0 & 27.8 & 48.1 \\
Qwen3.5-397B-A17B & 50.0 & 54.7 & 36.4 & 42.9 & 0.0 & 36.1 & 47.1 \\
Qwen3.5-9B & 31.8 & 32.8 & 27.3 & 28.6 & 0.0 & 19.4 & 28.9 \\
\midrule
\multicolumn{8}{>{\columncolor{piggypink!50}}l}{\emph{Specialized LLMs}\vspace{0.02in}} \\
NatureLM-8$\times$7B & 31.8 & 9.4 & 0.0 & 57.1 & 33.3 & 19.4 & 20.9 \\
SciReasoner-8B & 34.8 & 53.1 & 45.5 & 42.9 & 33.3 & 33.3 & 41.7 \\
TxGemma-9B & 7.6 & 14.1 & 0.0 & 57.1 & 100.0 & 16.7 & 14.4 \\
TxGemma-27B & 27.3 & 17.2 & 27.3 & 71.4 & 33.3 & 30.6 & 26.2 \\
\arrayrulecolor{black}\bottomrule
\end{tabular}
}
\vskip -5pt
\end{table}
\paragraph{Subtask-wise results for recognition stage.} The fine-grained performance across recognition subtasks is detailed in \cref{tab:recognition-rubric-breakdown-seq,tab:recognition-rubric-breakdown-struct,tab:recognition-rubric-breakdown-func} for sequence-level, structure-level, and domain/function recognition, respectively. The results reveal a clear hierarchy where most baselines excel at sequence-level tasks but struggle with higher-order structural and functional reasoning. Specifically, the strongest models reach 84.1--91.9 at the sequence level but at most 67.9 at the structure level and 71.1 on domain and function recognition. Secondary structure identification is the hardest subtask throughout, never exceeding 48.0 across all twelve models, and only GPT-5.4 and Gemini3.1-Pro exceed 70.0 on both family and superfamily classification. We also observe significant within-model variance, particularly in specialized architectures. For instance, TxGemma-9B attains a perfect pass rate on cellular component prediction yet 0.0 on molecular function, and NatureLM-8$\times$7B scores 0.0 on both motif detection and aromaticity prediction while achieving the highest disulfide bond identification rate of any model (80.0). These disparities suggest that specialized pretraining does not transfer uniformly to language-interfaced recognition.

\begin{table}[t]
\centering
\caption{Pass rates for each evaluation criterion within each engineering subtask across general-purpose and domain-specialized LLMs in tool-free setting.}
\label{tab:editing-rubric-breakdown}
\resizebox{\textwidth}{!}{
\begin{tabular}{l c c c c c c c c c c}
\toprule
& \multicolumn{3}{c}{Solubility} & \multicolumn{3}{c}{Stability} & \multicolumn{3}{c}{Activity} & \\
\cmidrule(lr){2-4} \cmidrule(lr){5-7} \cmidrule(lr){8-10}
Models & \makecell{Hard\\const.} & \makecell{Rat.\\align.} & \makecell{In silico\\valid.} & \makecell{Hard\\const.} & \makecell{Rat.\\align.} & \makecell{In silico\\valid.} & \makecell{Hard\\const.} & \makecell{Rat.\\align.} & \makecell{In silico\\valid.} & Overall \\
\midrule
\multicolumn{11}{>{\columncolor{piggypink!50}}l}{\emph{General-purpose LLMs}\vspace{0.02in}} \\
GPT-5.4 & {95.0} & {95.0} & {85.0} & {80.0} & {75.0} & {55.0} & {85.0} & {60.0} & {30.0} & \textbf{56.7} \\
Gemini3.1-Pro & 60.0 & 50.0 & 50.0 & 70.0 & {70.0} & {50.0} & 60.0 & 35.0 & 15.0 & 38.3 \\
Opus-4.6 & 0.0 & 0.0 & 0.0 & 15.0 & 15.0 & 10.0 & 0.0 & 0.0 & 0.0 & 3.3 \\
DeepSeek-V4-Pro & 65.0 & 50.0 & 50.0 & 60.0 & 55.0 & 40.0 & 75.0 & 40.0 & {30.0} & 40.0 \\
DeepSeek-V3.2 & 40.0 & 40.0 & 40.0 & {80.0} & 65.0 & 45.0 & 75.0 & {60.0} & {35.0} & 40.0 \\
Kimi-K2.5 & 5.0 & 0.0 & 0.0 & 10.0 & 10.0 & 5.0 & 5.0 & 0.0 & 0.0 & 1.7 \\
Qwen3.5-397B-A17B & 60.0 & {60.0} & {55.0} & 65.0 & 65.0 & {50.0} & 65.0 & {50.0} & {30.0} & \underline{45.0} \\
Qwen3.5-9B & 5.0 & 0.0 & 0.0 & 0.0 & 0.0 & 0.0 & 0.0 & 0.0 & 0.0 & 0.0 \\
\midrule
\multicolumn{11}{>{\columncolor{piggypink!50}}l}{\emph{Specialized LLMs}\vspace{0.02in}} \\
NatureLM-8$\times$7B & 70.0 & 0.0 & 0.0 & {90.0} & 10.0 & 10.0 & 60.0 & 0.0 & 0.0 & 3.3 \\
SciReasoner-8B & {90.0} & 0.0 & 0.0 & {90.0} & 0.0 & 0.0 & {95.0} & 15.0 & 10.0 & 3.3 \\
TxGemma-9B & 5.0 & 0.0 & 0.0 & 10.0 & 0.0 & 0.0 & 0.0 & 0.0 & 0.0 & 0.0 \\
TxGemma-27B & 15.0 & 0.0 & 0.0 & 5.0 & 0.0 & 0.0 & 20.0 & 0.0 & 0.0 & 0.0 \\
\arrayrulecolor{black}\bottomrule
\end{tabular}
}
\vskip -10pt
\end{table}
\paragraph{Subtask-wise results for engineering stage.} We decompose engineering stage into its subtasks and report their performance in \cref{tab:editing-rubric-breakdown}. According to the table, engineering performance is bottlenecked by rationale alignment and in silico validity rather than hard constraint compliance. The dissociation is most pronounced for specialized models: NatureLM-8$\times$7B and SciReasoner-8B satisfy hard constraints in 60.0--95.0\% of queries across all three subtasks, yet score zero on rationale alignment in two of the three subtasks each, namely solubility and activity for NatureLM-8$\times$7B and solubility and stability for SciReasoner-8B. General-purpose models instead degrade gradually, losing performance at each successive criterion rather than failing at a single one. Across both groups, activity is the hardest subtask, with in silico validity never exceeding 35.0 for any model.


\begin{table}[t]
\centering
\caption{Pass rates for each evaluation criterion within each GO-conditioned generation subtask across general-purpose and domain-specialized LLMs in tool-free setting.}
\label{tab:generation-rubric-breakdown-go}
\resizebox{\textwidth}{!}{
\begin{tabular}{l c c c c c c c c c c c c c}
\toprule
& \multicolumn{4}{c}{GO-MF} & \multicolumn{4}{c}{GO-MF \& GO-CC} & \multicolumn{4}{c}{GO-MF \& GO-BP} & \\
\cmidrule(lr){2-5} \cmidrule(lr){6-9} \cmidrule(lr){10-13}
Models & \makecell{Seq.\\valid.} & \makecell{Fold\\valid.} & \makecell{GO\\match} & Novelty & \makecell{Seq.\\valid.} & \makecell{Fold\\valid.} & \makecell{GO\\match} & Novelty & \makecell{Seq.\\valid.} & \makecell{Fold\\valid.} & \makecell{GO\\match} & Novelty & Overall \\
\midrule
\multicolumn{14}{>{\columncolor{piggypink!50}}l}{\emph{General-purpose LLMs}\vspace{0.02in}} \\
GPT-5.4 & 100.0 & 23.3 & 33.3 & 80.7 & 100.0 & 13.3 & 30.0 & 84.4 & 100.0 & 16.7 & 43.3 & 88.9 & 10.0 \\
Gemini3.1-Pro & 96.7 & 80.0 & 33.3 & 55.6 & 96.7 & 33.3 & 30.0 & 59.1 & 100.0 & 43.3 & 33.3 & 73.1 & 24.4 \\
Opus-4.6 & 100.0 & 26.7 & 26.7 & 77.4 & 100.0 & 13.3 & 30.0 & 78.3 & 100.0 & 0.0 & 33.3 & 91.0 & 6.7 \\
DeepSeek-V4-Pro & 100.0 & 40.0 & 16.7 & 82.9 & 100.0 & 30.0 & 36.7 & 80.6 & 100.0 & 43.3 & 43.3 & 86.9 & 17.8 \\
DeepSeek-V3.2 & 100.0 & 30.0 & 20.0 & 86.6 & 96.7 & 13.3 & 23.3 & 87.7 & 96.7 & 26.7 & 30.0 & 92.2 & 11.1 \\
Kimi-K2.5 & 100.0 & 13.3 & 13.3 & 85.8 & 100.0 & 10.0 & 30.0 & 90.1 & 100.0 & 10.0 & 23.3 & 93.9 & 10.0 \\
Qwen3.5-397B-A17B & 100.0 & 30.0 & 23.3 & 84.3 & 100.0 & 16.7 & 23.3 & 86.9 & 100.0 & 20.0 & 30.0 & 92.6 & 10.0 \\
Qwen3.5-9B & 100.0 & 6.7 & 0.0 & 97.5 & 100.0 & 20.0 & 3.3 & 98.1 & 100.0 & 0.0 & 10.0 & 98.5 & 0.0 \\
\midrule
\multicolumn{14}{>{\columncolor{piggypink!50}}l}{\emph{Specialized LLMs}\vspace{0.02in}} \\
NatureLM-8$\times$7B & 96.7 & 50.0 & 30.0 & 88.8 & 90.0 & 46.7 & 33.3 & 74.4 & 96.7 & 40.0 & 26.7 & 94.1 & \underline{27.8} \\
SciReasoner-8B & 100.0 & 60.0 & 53.3 & 86.3 & 100.0 & 46.7 & 43.3 & 88.1 & 96.7 & 23.3 & 33.3 & 90.4 & \textbf{30.0} \\
TxGemma-9B & 100.0 & 0.0 & 6.7 & 90.8 & 96.7 & 23.3 & 20.0 & 91.9 & 100.0 & 30.0 & 23.3 & 91.5 & 5.6 \\
TxGemma-27B & 80.0 & 6.7 & 3.3 & 90.8 & 86.7 & 6.7 & 13.3 & 92.0 & 93.3 & 6.7 & 20.0 & 91.7 & 2.2 \\
\arrayrulecolor{black}\bottomrule
\end{tabular}
}
\vskip -5pt
\end{table}
\begin{table}[t]
\centering
\caption{Pass rates for each evaluation criterion within each binder generation subtask across general-purpose and domain-specialized LLMs in tool-free setting.}
\label{tab:generation-rubric-breakdown-binder}
\resizebox{\textwidth}{!}{
\begin{tabular}{l c c c c c c c c c c c c c}
\toprule
& \multicolumn{3}{c}{Protein target} & \multicolumn{3}{c}{Small-molecule target} & \multicolumn{3}{c}{Protein target \& miniprotein} & \multicolumn{3}{c}{Protein target \& binding site} & \\
\cmidrule(lr){2-4} \cmidrule(lr){5-7} \cmidrule(lr){8-10} \cmidrule(lr){11-13}
Models & \makecell{Seq.\\valid.} & \makecell{Fold\\valid.} & \makecell{Interface\\valid.} & \makecell{Seq.\\valid.} & \makecell{Fold\\valid.} & \makecell{Interface\\valid.} & \makecell{Seq.\\valid.} & \makecell{Fold\\valid.} & \makecell{Interface\\valid.} & \makecell{Seq.\\valid.} & \makecell{Fold\\valid.} & \makecell{Interface\\valid.} & Overall \\
\midrule
\multicolumn{14}{>{\columncolor{piggypink!50}}l}{\emph{General-purpose LLMs}\vspace{0.02in}} \\
GPT-5.4 & 90.0 & 80.0 & 3.3 & 96.7 & 73.3 & 20.0 & 83.3 & 86.7 & 0.0 & 100.0 & 90.0 & 0.0 & 5.8 \\
Gemini3.1-Pro & 100.0 & 86.7 & 3.3 & 100.0 & 90.0 & 26.7 & 100.0 & 93.3 & 0.0 & 100.0 & 96.7 & 0.0 & \textbf{7.5} \\
Opus-4.6 & 100.0 & 53.3 & 0.0 & 100.0 & 16.7 & 3.3 & 96.7 & 90.0 & 0.0 & 100.0 & 70.0 & 0.0 & 0.0 \\
DeepSeek-V4-Pro & 100.0 & 83.3 & 6.7 & 100.0 & 76.7 & 20.0 & 96.7 & 93.3 & 0.0 & 100.0 & 93.3 & 0.0 & \underline{6.7} \\
DeepSeek-V3.2 & 100.0 & 90.0 & 0.0 & 100.0 & 86.7 & 3.3 & 96.7 & 90.0 & 3.3 & 100.0 & 93.3 & 0.0 & 1.7 \\
Kimi-K2.5 & 100.0 & 53.3 & 3.3 & 100.0 & 46.7 & 3.3 & 66.7 & 83.3 & 0.0 & 96.7 & 56.7 & 0.0 & 0.8 \\
Qwen3.5-397B-A17B & 100.0 & 83.3 & 0.0 & 100.0 & 66.7 & 10.0 & 96.7 & 93.3 & 0.0 & 100.0 & 86.7 & 0.0 & 2.5 \\
Qwen3.5-9B & 86.7 & 70.0 & 6.7 & 96.7 & 80.0 & 0.0 & 16.7 & 56.7 & 6.7 & 80.0 & 70.0 & 16.7 & 5.0 \\
\midrule
\multicolumn{14}{>{\columncolor{piggypink!50}}l}{\emph{Specialized LLMs}\vspace{0.02in}} \\
NatureLM-8$\times$7B & 100.0 & 70.0 & 16.7 & 100.0 & 16.7 & 0.0 & 93.3 & 93.3 & 0.0 & 100.0 & 70.0 & 10.0 & \underline{6.7} \\
SciReasoner-8B & 96.7 & 36.7 & 0.0 & 96.7 & 16.7 & 3.3 & 100.0 & 86.7 & 3.3 & 100.0 & 53.3 & 0.0 & 1.7 \\
TxGemma-9B & 100.0 & 83.3 & 0.0 & 100.0 & 36.7 & 10.0 & 100.0 & 86.7 & 0.0 & 100.0 & 83.3 & 0.0 & 2.5 \\
TxGemma-27B & 100.0 & 70.0 & 0.0 & 100.0 & 3.3 & 0.0 & 96.7 & 90.0 & 0.0 & 96.7 & 76.7 & 0.0 & 0.0 \\
\arrayrulecolor{black}\bottomrule
\end{tabular}
}
\vskip -12pt
\end{table}
\paragraph{Subtask-wise results for generation stage.} We provide a detailed breakdown of generation performance across GO term-conditioned subtasks in \cref{tab:generation-rubric-breakdown-go} and binder design subtasks in \cref{tab:generation-rubric-breakdown-binder}. Across both generation tracks, performance is bottlenecked by functional grounding
rather than by global sequence plausibility. Sequence validity and novelty remain high for nearly every model, yet GO match never exceeds 53.3 in GO-conditioned generation and interface validity never exceeds 26.7 in binder generation. The two bottlenecks respond differently to additional conditions, where adding a second GO term leaves GO match essentially unchanged and in some models slightly higher, whereas imposing a miniprotein length or binding-site constraint drives interface validity to 0.0 for nine and ten of the
twelve models, respectively. The difficulty therefore lies in grounding function for GO-conditioned generation, and in meeting explicitly stated geometric specifications for binder generation. 

\begin{table}[t]
\centering
\small
\caption{%
Calibration of in silico evaluators against experimental binding and stability
measurements and annotated GO conditions. All values are percentages. Bold denotes
the best value within each panel. Accuracy denotes the accuracy in predicting property direction relative to the wild-type (WT). Balanced accuracy (BA), shaded, averages the two
class-specific recalls.%
}
\label{tab:evaluator-reliability}
\begingroup
\setlength{\tabcolsep}{4pt}
\renewcommand{\arraystretch}{1.05}
\setlength{\aboverulesep}{0pt}\setlength{\belowrulesep}{0pt}
\definecolor{bacolumn}{HTML}{F7EEF1}
\newcommand{\panel}[1]{\rule{0pt}{2.5ex}\textbf{#1}}
\newcommand{\metric}[1]{\textnormal{#1}}
\newcommand{\panelgap}{\addlinespace[3.5pt]\midrule\addlinespace[1.5pt]}
\begin{tabular}{@{}
  >{\raggedright\arraybackslash}p{\dimexpr\linewidth-44.5mm-8\tabcolsep\relax}
  >{\centering\arraybackslash}p{11mm}
  >{\centering\arraybackslash}p{9.5mm}
  >{\centering\arraybackslash}p{16mm}
  >{\columncolor{piggypink!50}\centering\arraybackslash}p{8mm}@{}}
\toprule
\panel{A\quad Binder screening}
  & \metric{Precision} & \metric{Recall} & \metric{Specificity} & \metric{BA} \\
\cmidrule[0.4pt]{1-5}
Protenix $\mathrm{ipTM}>0.5$ & 46.1 & \textbf{95.6} & 42.4 & 69.0 \\
Protenix $\mathrm{ipTM}>0.8$ & \textbf{58.7} & 94.1 & \textbf{65.9} & \textbf{80.0} \\
$\mathrm{pDockQ}>0.23$       & 57.0 & \textbf{95.6} & 62.9 & 79.2 \\
Rosetta $S_c\geq0.55$        & 50.9 & 82.4 & 59.1 & 70.7 \\
\panelgap
\panel{B\quad Stability-change prediction}
  & & & \metric{Accuracy} & \metric{BA} \\
\cmidrule[0.4pt]{1-5}
PyRosetta $\Delta\Delta G$       & & & 56.0 & 55.8 \\
ThermoMPNN & & & 65.0 & 59.0 \\
MGnify stability predictor            & & & \textbf{74.5} & \textbf{72.5} \\
\panelgap
\panel{C\quad Ligand-affinity change prediction}
  & & & \metric{Accuracy} & \metric{BA} \\
\cmidrule[0.4pt]{1-5}
Vina (Protenix-v1 WT structure + Rosetta relax)                  & & & 56.4 & 49.8 \\
Vina (Protenix-v1 WT/mutant structures)                  & & & 60.8 & 58.3 \\
Boltz-2 affinity (Protenix-v1 WT/mutant structures)         & & & 66.3 & 56.3 \\
Boltz-2 affinity (Boltz-2 WT/mutant structures)             & & & \textbf{66.9} & \textbf{61.5} \\
\panelgap
\panel{D\quad GO-condition consistency} &
  & \metric{Precision} & \metric{Recall} & \metric{BA} \\
\cmidrule[0.4pt]{1-5}
ESM-C GO-reference matching & & 64.5 & \textbf{73.3} & \textbf{66.5} \\
DeepGO-SE         & & \textbf{93.8} & 27.6 & 62.9 \\
GO-GPT                     & & 89.2 & 14.9 & 56.6 \\
\bottomrule
\end{tabular}
\endgroup
\vskip -5pt
\end{table}
\paragraph{Reliability of in silico evaluators.}
We select each in silico evaluator by calibrating candidate configurations against experimental measurements, drawing on thresholds and predictors established in prior protein-design studies (\cref{tab:evaluator-reliability}).

For binder screening, the $\mathrm{ipTM} > 0.8$ threshold has been adopted by PXDesign alongside pTM and structural-consistency filters~\citep{pxdesign}. Additionally, on 200 experimentally characterized binder--target pairs spanning four targets~\citep{bennett2023improving}, we compare this cutoff against $\mathrm{ipTM}>0.5$ and the published cutoffs for pDockQ~\citep{burke2023network} and Rosetta shape
complementarity~\citep{lawrence1993shape,bindcraft}. As shown, $\mathrm{ipTM} > 0.8$ threshold attains the highest precision, specificity, and balanced accuracy (BA; $80.0\%$) while retaining $94.1\%$ recall.

For stability, the MGnify absolute stability predictor is the most reliable of the three estimators we examined. We adopt its released augmented SaProt$\Delta G$ ensemble~\citep{cho2026accurate} and evaluate 200 WT--single-mutant pairs from its official homology-filtered MGnify test split, further excluding homologs of ThermoMPNN training proteins by $30\%$ sequence-identity screening and JackHMMER. Given independently predicted Protenix-v1 WT and mutant structures with MSAs enabled, it classifies the direction of stability change with $74.5\%$ accuracy and $72.5\%$ BA, ahead of ThermoMPNN and of unrelaxed PyRosetta
energy differences.

For ligand affinity, the native Boltz-2 structure-and-affinity pipeline is the strongest configuration we tested. We predict WT--ligand and mutant--ligand complexes independently with MSAs enabled~\citep{boltz2} on 181 paired Platinum records~\citep{pires2015platinum}. This recovers the direction of affinity change with $66.9\%$ accuracy and $61.5\%$ BA, against $60.8\%$/$58.3\%$ for Vina scored on the same Protenix-v1 structures. 

For GO-conditioned generation, embedding-based reference matching gives the best balance between accepting annotated conditions and rejecting mismatched ones. We compare DeepGO-SE's ontology-aware sequence classifier~\citep{deepgo-se}, GO-GPT's autoregressive GO-term predictor~\citep{BioReason-Pro}, and our \emph{ESM-C GO-reference matching} rule, which accepts a sequence when its minimum cosine similarity to ten proteins sharing all requested GO labels is at least the minimum similarity between 10 distinct reference proteins; MF+BP and MF+CC additionally require both labels to be recovered for the same protein. This choice of representation is supported by prior work showing that ESM-C embeddings track experimentally determined functional equivalence more faithfully than sequence identity~\citep{shaw2025evaluating}. Measuring BA as the average of annotated-condition acceptance and mismatched-condition rejection, the resulting rule attains the highest recall ($73.3\%$) and BA ($66.5\%$); the classifier baselines are considerably more precise ($93.8\%$ for DeepGO-SE) but recover only $27.6\%$ and $14.9\%$ of annotated conditions, which would leave most functionally valid designs unscored.

\begin{figure}[t!]
    \centering
    \includegraphics[width=\linewidth]{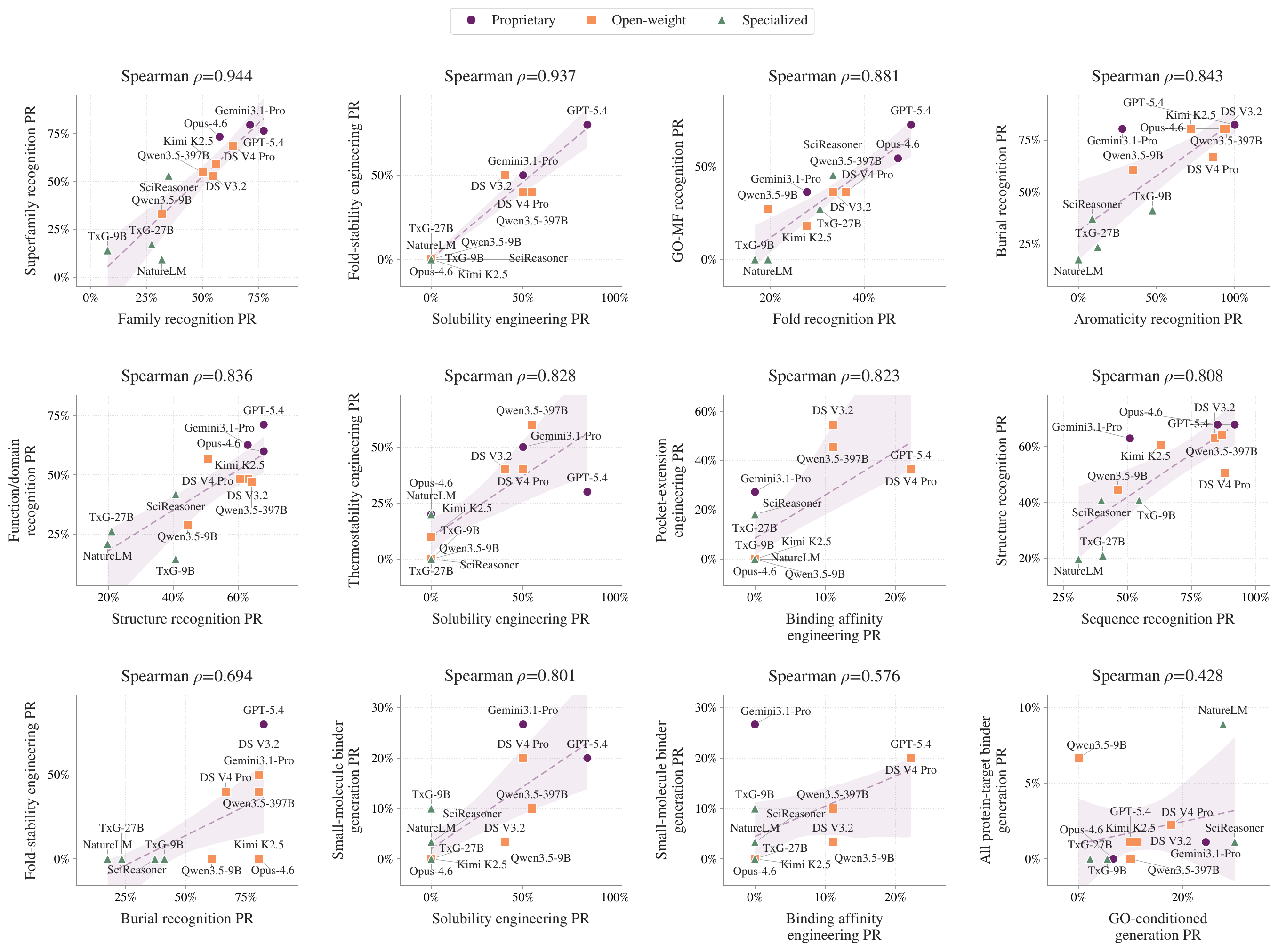}
    \caption{
    Correlations between different subtask performance of baseline LLMs. DS denotes DeepSeek and TxG indicates TxGemma-chat models.
    }
    \label{fig:task-correlation-appendix}
    \vskip -10pt
\end{figure}

\paragraph{Cross-subtask correlations.}
To examine our coherence claim more broadly than the three representative pairs reported in the main paper, we measure pairwise Spearman correlations across diverse mechanistically linked subtask pairs in \system{} (\cref{fig:task-correlation-appendix}). The twelve pairs span recognition$\times$recognition, engineering$\times$engineering, recognition$\times$engineering, engineering$\times$generation, and generation$\times$generation combinations. All of them correlate positively and nine reach $\rho\geq0.80$, with the strongest coupling found within stages: family$\times$ superfamily recognition ($\rho=0.944$), solubility $\times$ fold-stability engineering ($\rho=0.937$), fold $\times$ GO-MF recognition ($\rho=0.881$), and binding-affinity $\times$ pocket-extension engineering ($\rho=0.823$). The weakest pair is GO-conditioned $\times$ protein-target binder generation ($\rho=0.428$), which is expected given that generation pass rates for protein binders are compressed near zero for most models and therefore admit little rank separation. The breadth of these positive correlations shows that the coupling reported in the main paper is not an artifact of the three pairs we chose to highlight, further supporting that recognition, engineering, and generation probe mutually dependent rather than independent competences, and are therefore evaluated together.

\begin{table}[t]
\centering
\small
\caption{%
Safety refusal rates (\%) on our benchmark. Each entry gives the percentage of queries that
returned no usable answer, with the raw count in parentheses. API-level safety
blocks and explicit in-text refusals are counted together, and all such cases are
scored as failures.%
}
\label{tab:safety-refusals}
\begingroup
\setlength{\tabcolsep}{7pt}
\renewcommand{\arraystretch}{1.12}
\begin{tabular}{@{}l
  >{\centering\arraybackslash}p{22mm}
  >{\centering\arraybackslash}p{22mm}
  >{\centering\arraybackslash}p{22mm}@{}}
\toprule
Models & Recognition & Engineering & Generation \\
\midrule
Opus-4.6 (w/o tool) & 0.0 & 35.0 & 0.0 \\
GPT-5.4 (w/o tool)  & 0.0 & \pz3.3 & 7.1 \\
GPT-5.4 (w/ tool; ReAct-based)   & 0.0 & \pz3.3 & 1.0 \\
\bottomrule
\end{tabular}
\endgroup
\vskip -10pt
\end{table}
\noindent\textbf{Safety refusals.} A fraction of queries returned no usable answer because of safety filtering, and we score every such case as a failure (\cref{tab:safety-refusals}). We report the refusal rate per stage, counting API-level safety blocks together with explicit in-text refusals. Refusals are absent in recognition and concentrate in engineering for Opus-4.6, which declined 35.0\% of its engineering queries.

\section{Protein Recognition Task Details}\label{C}

\subsection{Sequence-level recognition} 
\begin{itemize}[topsep=0pt,itemsep=1mm, parsep=0pt, leftmargin=5mm]
    \item \textbf{Motif detection:} This task probes whether the model can interpret biologically meaningful sequence patterns rather than recognizing individual residues in isolation. Each question presents a protein sequence with a PROSITE~\citep{prosite} sequence motif pattern and asks the model to determine whether the motif occurs in the sequence and, if so, to report all matching residue spans using one-indexed positions. 

    \item \textbf{Physicochemical composition:} Given a protein sequence, the model predicts global biochemical properties of the entire sequence. We include isoelectric point (pI), hydropathy as measured by GRAVY score, and aromaticity, using sequence-property calculations implemented in BioPython's \texttt{ProteinAnalysis} module~\citep{biopython}. We formulate pI and GRAVY as categorical recognition over predefined biochemical ranges, and query aromaticity as the numeric fraction of aromatic residues.

    \item \textbf{Residue-level property recognition:} To test residue-level understanding, we sample individual one-indexed residue positions and ask the model to classify their biochemical traits. Specifically, the model identifies net charge of each residue as positive, negative, or neutral, and its hydropathy as hydrophobic or hydrophilic. 
\end{itemize}

\subsection{Structure-level recognition} 
To construct queries in structure-level recognition task, we use experimentally determined 3D structures from reference sequences from the RCSB PDB database~\citep{RCSB_PDB}.

\begin{itemize}[topsep=0pt,itemsep=1mm, parsep=0pt, leftmargin=5mm]
\item \textbf{Secondary structure identification:} Each question asks the model to identify which candidate subsequence corresponds to a specified secondary-structure class, such as helix, sheet, or coil. We assign secondary-structure labels from the associated PDB structure using DSSP, and sample length-matched distractors from subsequences assigned to different structural classes.

\item \textbf{Burial classification:} To evaluate structure-aware reasoning about solvent exposure, we ask the model to identify a subsequence that is either buried or exposed. We obtain residue-level burial labels from the DSSP-computed relative solvent accessibility (RSA), applying fixed thresholds to distinguish buried and surface-exposed residues. We then sample length-matched distractors from regions of the opposite burial class.

\item \textbf{Disulfide bond identification:} Each question asks the model to identify which residue pair forms a disulfide bond in the associated structure. Correct answers are Cys--Cys pairs confirmed by structural records or geometry, while distractors include both non-bonded Cys--Cys pairs and Cys--non-Cys pairs, to prevent the model from solving the task by cysteine frequency alone.
\end{itemize}

\subsection{Domain/function recognition} 
\begin{itemize}[topsep=0pt,itemsep=1mm, parsep=0pt, leftmargin=5mm]
    \item \textbf{Fold recognition:} Each question presents a reference protein sequence and asks the model to identify another sequence that shares the same topology group (T-group) in the ECOD database~\citep{ecod}, which classifies protein domains by their evolutionary relationships. To prevent the model from shortcutting through sequence homology, we constrain the correct answer to share the same T-group as the reference yet fall below 30\% sequence identity.
    
    \item \textbf{Family/Superfamily classification:} Given a protein sequence, the model identifies the correct family and superfamily labels, which we curate from the InterPro database~\citep{interpro} -- a resource that integrates diverse protein signatures into unified functional and evolutionary classifications. To construct hard negatives that are structurally similar yet functionally distinct, we draw distractors from the same superfamily as the reference for family classification, and from the same ECOD X-group~\citep{ecod} for superfamily classification, a category that groups proteins with possible but inconclusive evolutionary relationships.
    
    \item \textbf{GO term identification:} To probe function-level recognition, we ask the model to identify the correct Gene Ontology (GO)~\citep{go_benchmark} annotations for a given protein sequence. For each reference protein, we select the top-3 leaf terms whose associated Swiss-Prot entries appear least frequently in the database, targeting specific rather than generic functional roles that would likely overlap with other proteins. We then draw distractors from other Swiss-Prot proteins that share the same superfamily as the reference but whose top-3 least frequent GO terms do not overlap with any of its curated terms.
\end{itemize}


\section{Rationale-guided Engineering Task Details}\label{D}
\subsection{Expert-curated rationales}

For each subtask, a domain expert encodes the design rationale as a structured pair of rules: a family of defective patterns that identifies residues $M$ whose modification is expected to improve the target property, and a family of residues $P$ that must be preserved to avoid disrupting function, fold, or oligomeric context. Every per-residue call is backed by a computationally grounded feature--including residue exposure, local structure, ligand geometry, residue chemistry, functional annotations, or group-wise evolutionary statistics--so that the rationale is inspectable rather than opaque. We provide the full rule sets below.

\paragraph{Solubility.}
The solubility rationale targets solvent-exposed chemical liabilities. A residue enters the defective pool if it is surface exposed, defined by relative SASA $\mathrm{RSA} > 0.25$, and is hydrophobic either by amino-acid class or Kyte--Doolittle hydropathy ($>0$). These residues are assigned a charged-or-polar repair direction when homologous sequences support such substitutions. We also include exposed unpaired cysteines: a Cys with $\mathrm{RSA} > 0.25$ and no detected disulfide partner is treated as a liability and assigned a Cys$\rightarrow$Ser repair direction.

The preserved pool contains residues whose mutation may compromise fold or function: disulfide-bonded cysteines, ligand/cofactor contacts, salt-bridge participants, oligomeric interface residues, annotated catalytic or active-site residues, annotated binding-site residues, buried hydrophobic core residues, highly conserved surface residues lacking polar/charged substitution support, and Pro/Gly residues with potential backbone-geometry roles. Structural contacts are computed from the input structure, while catalytic, active-site, binding-site, and disulfide annotations are also imported from UniProt when available.

\paragraph{Fold Stability.}
The fold-stability rationale targets local structural weaknesses with pattern-specific repair directions. We identify under-packed core sites as buried small hydrophobic residues ($\mathrm{RSA}<0.09$) with low local C$_\beta$-neighbor packing density; these are assigned larger hydrophobic substitutions. Buried polar residues without a side-chain hydrogen-bond partner are assigned hydrophobic-isostere substitutions. Helix-related defects include Gly in helix interiors, suboptimal helix N-caps, and helix N-terminal dipole positions lacking acidic stabilization, with repairs such as Gly$\rightarrow$Ala, N-cap-capable residues, or Asp/Glu. We also include loop positions with Pro-compatible $\phi$ angles, buried unpaired charged residues, and positions where the wild type is rare relative to a high-frequency consensus residue.

Protected residues for fold stability include catalytic or active-site residues, metal-coordinating residues, disulfide-bonded cysteines, ligand/cofactor contacts, oligomeric interface residues, buried highly conserved core residues, salt-bridge participants, Pro/Gly residues, and residues participating in satisfied side-chain hydrogen-bond networks in densely packed neighborhoods. 

\paragraph{Thermostability.}
The thermostability rationale is based on mesophile--thermophile group contrasts. For each mesophilic target, homologs are partitioned into mesophilic and thermophilic groups using organism-level optimal growth temperature metadata. A residue is considered defective when the mesophilic wild-type amino acid is common in the mesophile group but depleted in the thermophile group, with a thermophile-enriched non-wild-type amino acid providing the suggested substitution. In the implemented rule, the mesophile wild-type frequency must be at least $0.4$, the thermophile wild-type frequency at most $0.3$, the mesophile--thermophile frequency gap at least $0.2$, and the thermophile consensus frequency at least $0.2$.

The preserved pool contains catalytic or active-site residues, metal-proximal residues, disulfide-bonded cysteines, ligand/cofactor-proximal residues, interface residues, residues conserved in both temperature groups, Pro/Gly residues, and salt-bridge participants when detected. 

\paragraph{Activity--binding affinity.}
The binding-affinity rationale targets chemical-complementarity mismatches at the substrate-binding site. The pipeline selects a substrate ligand from the structure, classifies ligand atoms by local chemistry, and computes the closest substrate heavy atom for each protein residue. Residues are binned as contacting, near, or far using minimum heavy-atom distance thresholds of $4.5$ \AA{} and $6.5$ \AA{}. A contacting residue enters the defective pool when its residue class mismatches the nearest substrate atom chemistry: hydrophobic residues facing polar atoms, hydrophobic residues facing charged groups, polar residues facing nonpolar or aromatic atoms, or same-sign charge repulsion. The repair direction is the complementary residue class, such as polar, hydrophobic, or the opposite charge.

Protected residues include annotated catalytic or active-site residues, metal-coordinating residues, disulfide-bonded cysteines, already complementary substrate-contacting residues, highly conserved substrate-contacting residues, non-binding residues far from the substrate, interface residues, Pro/Gly residues near the binding site, and salt-bridge participants. If a residue is both a candidate mismatch and protected, the protection rule takes precedence.

\paragraph{Activity--pocket expansion.}
The pocket-expansion rationale targets steric restriction in the ligand-binding pocket. Pocket membership is defined directly from ligand geometry: residues with minimum heavy-atom distance to the ligand at most $6.0$ \AA{} are pocket-lining residues, residues within $8.0$ \AA{} are pocket-adjacent, and all others are non-pocket residues. Each residue is assigned a side-chain volume and size class; residues with volume below $120$ \AA$^3$ are small, $120$--$170$ \AA$^3$ are medium, and at least $170$ \AA$^3$ are large. A defective pocket-expansion target must be a medium or large pocket-lining residue with homologous support for smaller substitutions. Candidate substitutions are amino acids whose side-chain volume is at least $10$ \AA$^3$ smaller than the wild type.

The preserved pool includes catalytic residues, metal-coordinating residues, disulfide-bonded cysteines, essential substrate contacts, conserved pocket residues lacking smaller-residue support, non-pocket residues, interface residues, Pro/Gly residues in or near the pocket, and salt-bridge participants. Pocket expansion is therefore defined by ligand-distance bins and residue volume.

\subsection{Answer evaluation}

All proposed mutants are evaluated through a multi-facet pipeline that proceeds sequentially: hard constraints, then rationale alignment, and finally in silico validation. We adopt this dependent structure because each preceding rubric establishes a precondition for the next to be meaningful. Mutating a protected residue, such as a catalytic residue or a binding-site contact, can disrupt the very function the engineering task aims to preserve, so a property gain achieved at such a cost cannot be credited as successful engineering. Likewise, even when the hard constraints are satisfied, a property gain that does not follow the expert-specified rationale is achieved by a mechanism unrelated to the intended design intent, and therefore does not reflect the rationale-guided competence the task is meant to probe. We accordingly skip downstream rubrics for mutants that fail an earlier one. A prediction is counted as correct only when it passes every applicable rubric.

\paragraph{Hard constraints.} This rubric verifies the prerequisite conditions a mutant must satisfy to qualify for further in silico evaluation.
\begin{itemize}[topsep=0pt,itemsep=1mm, parsep=0pt, leftmargin=5mm]
    \item Mutant sequence length must be identical to the WT sequence.
    \item Mutant sequence must not be identical to the WT sequence.
    \item Protected positions must remain unchanged.
    \item The number of mutations must not exceed the predefined maximum.
    \item Mutations are allowed only at specified target positions.
    \item Non-target and non-mutated positions must remain WT.
\end{itemize}

\paragraph{In silico validation.}
This rubric combines a common fold-quality gate with a task-specific property metric. The common pLDDT gate passes when $\text{pLDDT}_{\text{mutant}} \geq \text{pLDDT}_{\text{WT}}$ or $\text{pLDDT}_{\text{mutant}} > 70$.

\begin{itemize}[topsep=0pt,itemsep=1mm, parsep=0pt, leftmargin=5mm]
    \item \textbf{Solubility:}
    The mutant must pass the pLDDT gate from Protenix-v1, decrease surface GRAVY, and increase the fraction of surface residues that are charged or polar. Surface residues are recomputed from predicted structures using RSA $>0.25$.

    \item \textbf{Fold stability and thermostability:}
    The mutant must pass the pLDDT gate from Protenix-v1 and increase the unfolding free energy $\Delta G_{\mathrm{unfold}}$ predicted from MGnify stability predictor.
    We define
    $\Delta\Delta G_{\mathrm{unfold}}
    = \Delta G_{\mathrm{unfold}}^{\mathrm{mut}}
    - \Delta G_{\mathrm{unfold}}^{\mathrm{WT}}$;
    thus, $\Delta\Delta G_{\mathrm{unfold}} > 0$ indicates increased stability.

    \item \textbf{Binding affinity:}
    The mutant must pass the pLDDT gate and have a lower
    $\log_{10}(\mathrm{IC}_{50}/\mu\mathrm{M})$
    than WT for the same ligand from the Boltz-2 prediction. Thus, a lower value indicates stronger predicted binding.

    \item \textbf{Pocket expansion:}
    The mutant must pass the pLDDT gate from Protenix-v1 and increase or preserve the active-site pocket volume measured by pyKVFinder:
    $\mathrm{V}^{\text{pk}}_{\mathrm{mutant}} \geq \mathrm{V}^{\text{pk}}_{\mathrm{WT}}$.
    The active-site cavity is selected using ligand coordinates and projected pocket residues when available.
\end{itemize}

\paragraph{Rationale alignment.}
For queries that include defective targets $M$, we additionally assess whether each mutated target follows the intended repair rule. 

\begin{itemize}[topsep=0pt,itemsep=1mm, parsep=0pt, leftmargin=5mm]
    \item \textbf{Solubility:}
    Surface-hydrophobic targets must be changed to an allowed charged or polar residue, excluding cysteine. Free-surface cysteine targets receive credit for Cys$\rightarrow$Ser.

    \item \textbf{Fold stability:}
    Mutations must satisfy the pattern-specific repair rule: larger hydrophobics for under-packed core sites, hydrophobic replacements for buried unsatisfied polar residues, helix-compatible substitutions for helix defects, Pro for Pro-compatible loops, neutral nonpolar replacements for buried unpaired charges, or the consensus/same-class residue for back-to-consensus targets.

    \item \textbf{Thermostability:}
    Mutations must be thermophile-directed: the mutant residue must match the suggested thermophile-enriched residue, belong to the acceptable thermophile-dominant amino-acid set, or satisfy an explicitly specified acceptable class.

    \item \textbf{Binding affinity:}
    Mutations must improve residue--substrate complementarity by matching the ideal residue character for the detected mismatch, such as polar, hydrophobic, or opposite charge.

    \item \textbf{Pocket expansion:}
    Pocket-lining targets must be replaced by an accepted smaller residue, or by any amino acid whose side-chain volume satisfies
    $\mathrm{V}^{\text{sc}}_{\mathrm{mutant}} \leq \mathrm{V}^{\text{sc}}_{\mathrm{WT}} - \SI{10}{\angstrom^3}$.
\end{itemize}




\paragraph{Search terms for literature-based filtering.}

We first built task-specific literature pools from PubMed and Semantic Scholar using broad engineering queries, and then filtered out a candidate if a returned title or abstract contained the protein name together with at least one design term and at least one task-context term.

\begin{itemize}[topsep=0pt,itemsep=0.5mm,parsep=0pt,leftmargin=5mm]

\item \textbf{Solubility and fold stability.}
We used broad protein-engineering queries covering directed evolution, de novo/computational/rational design, mutation, variants, and engineering. Design terms included protein engineering/editing, computational or rational design, mutation, and solubility/stability/thermostability improvement; context terms included protein engineering, solubility, stability, and thermostability.

\item \textbf{Activity.}
We used enzyme-focused queries covering enzyme engineering, catalytic activity, mutation, $k_{\operatorname{cat}}$, and directed evolution. Design terms included enzyme/protein editing or mutation, catalytic improvement, affinity enhancement, interaction engineering, and enzyme--substrate or enzyme--inhibitor binding; context terms included enzyme, protein engineering, and activity.

\item \textbf{Thermostability.}
We used temperature-focused queries covering thermostability improvement, thermal stability, thermophile/mesophile contrasts, protein mutation, and melting temperature. Design terms included mesophile/thermophile engineering or mutation, protein engineering/editing, computational or rational design, and thermostability improvement; context terms included thermostability, thermal stability, Tm, heat resistance, thermophile/mesophile, thermotolerance, thermal denaturation, thermolability, and optimal growth temperature.

\end{itemize}

\section{Generation Task Details}\label{E}

\subsection{Answer evaluation}

We evaluate generated protein sequences along several axes. We first check sequence validity, requiring a non-empty sequence composed only of canonical amino acid letters. For miniprotein-conditioned generation, the sequence must additionally fall within the 40--80 amino acid length range.

For coarse-grained functional generation, we evaluate structural plausibility and functional consistency. Structural plausibility is assessed from predicted structure confidence, requiring $\text{pLDDT} >70$ and $\text{pTM} > 0.5$. Functional consistency is assessed with an ESM-C-based GO-reference matching rule~\citep{esmfold2}. 

For target-specific binder generation, we evaluate structural plausibility and target engagement. We require a valid sequence, $\text{pLDDT} >70$, $\text{pTM} > 0.5$, and $\text{ipTM} >0.8$~\citep{bindcraft, bindenergycraft}. Under the binding-site-conditioned setting, we further require the binder to contact at least half of the specified target-site residues within a 5~\AA{} heavy atom distance cutoff. 

We additionally report sequence novelty across both GO-conditioned and binder-design queries. For each scorable GO-conditioned output, novelty is defined as 100 times one minus its maximum ungapped sequence identity to experimentally annotated Swiss-Prot proteins satisfying all queried GO terms. For each scorable binder, novelty is defined as 100 times one minus its maximum aligned sequence identity among significant MMseqs2 hits (E-value at most 1) in UniRef50 release (2026-03); a completed search with no hit receives a score of 100. We report the query-weighted mean over all scorable GO and binder outputs, rather than an equal-weight average of the two cohort means. 

The overall generation score is a strict pass: a query is counted as correct only when the sequence passes every required check for its subtask.

\subsection{Source dataset curation and contamination mitigation}

We curate three categories of source data for the generation stage, namely functional GO terms, protein targets, and small-molecule targets. The integrity of the binder track relies on a strict temporal holdout (2025-09-01) for curating protein binding targets from UniProt and small-molecule targets from PubChem entries. This ensures that the structural or chemical targets are ``novel'' to the model. 

\begin{itemize}[topsep=0pt,itemsep=1mm, parsep=0pt, leftmargin=5mm]
    \item \textbf{GO functional terms:} We curate GO terms in the way that balances functional rarity. Specifically, we compute the information content (IC)~\citep{go_benchmark} for each GO term combination and partition the candidate pool into equal-mass quantile bins. Taking MF--BP paired conditioning as an example, the IC of each pair $(\text{MF}_i,\text{BP}_j)$ with MF term $i$ and BP term $j$ is $\text{IC}(\text{MF}_i,\text{BP}_j)=-\log_2\frac{|\mathcal P(\text{MF}_i)\cap\mathcal P(\text{BP}_j)|}{|\mathcal P|}$, where $\mathcal P(\cdot)$ denotes the set of Swiss-Prot proteins annotated with given GO term and $\mathcal P$ is the full Swiss-Prot collection. 
    \item \textbf{Protein targets:} We curate high-resolution ($\leq$ 3.0 \AA) dimeric structures from the RCSB PDB~\citep{RCSB_PDB}. Targets are restricted to those where at least one chain was registered after the September 1, 2025 cutoff, ensuring they were not present in the training sets of current LLMs. We exclude enzymes and DNA/RNA-binding complexes to maintain track-specific focus. For site-specific subtasks, we identify the top-10 residues on the target chain with the highest density of heavy-atom contacts ($\leq$ 5 \AA) to the partner chain.
    \item \textbf{Small-molecule targets:} Ligands are sourced from PubChem~\citep{PubChem2023} using a creation-date cutoff of September 1, 2025. We apply a stringent artifact filter based on the Plinder badlist to remove buffers, reagents, and other non-biological molecules. SMILES strings are canonicalized using RDKit, and targets with extreme charges or long hydrocarbon linkers are discarded to ensure chemical relevance.
\end{itemize}


\section{Extended Comparison with Existing Benchmarks}\label{F}

\begin{table}[t]
\centering
\small
\setlength{\tabcolsep}{4pt}
\renewcommand{\arraystretch}{1}
\caption{Comparison of \system{} with existing protein-design benchmarks. \emph{Rec.}: recognition; \emph{Eng.}: engineering; \emph{Gen.}: generation. \emph{Lang.\ interface}: natural-language queries vs.\ rigid input schemas. \emph{Mech.\ rationale}: expose expert mechanistic rationales during sequence design. \emph{Struct.\ quality}: folded-structure validation beyond sequence-level metrics. \emph{Contam.\ control}: explicit mitigation via temporal holdout, structure/sequence split, or literature filtering.}
\label{tab:bench-comparison}
\begin{tabular}{lccccccc}
\toprule
Benchmark & \makecell{Lang.\\ interface} & \makecell{Rec.\\stage} & \makecell{Eng.\\stage} & \makecell{Gen.\\stage} & \makecell{Mech.\\ rationale} & \makecell{Struct.\\ quality} & \makecell{Contam.\\ control} \\
\midrule
PDB-Struct~\citep{Pdb-struct}                     & \xmark & \xmark & \xmark & \cmark & \xmark & \cmark & \cmark \\
ProteinBench~\citep{proteinbench}            & \xmark & \xmark & \xmark & \cmark & \xmark & \cmark & \cmark \\
ProteinLMBench~\citep{ProteinLMBench}      & \cmark & \cmark & \xmark & \xmark & \xmark & \xmark & \xmark \\
LiveProteinBench~\citep{liveproteinbench}  & \cmark & \cmark & \xmark & \xmark & \xmark & \xmark & \cmark \\
PDFBench~\citep{pdfbench}                 & \cmark & \xmark & \xmark & \cmark & \xmark & \cmark & \cmark \\
InstructProBench~\citep{instructpro}       & \cmark & \xmark & \xmark & \cmark & \xmark & \cmark & \cmark \\
Mol-Instructions~\citep{Mol-instructions}               & \cmark & \cmark & \xmark & \cmark & \xmark & \xmark & \xmark \\
SciReasoner~\citep{scireasoner}            & \cmark & \cmark & \xmark & \cmark & \xmark & \xmark & \xmark \\
\midrule
\rowcolor{piggypink}\system{} (Ours)                        & \cmark & \cmark & \cmark & \cmark & \cmark & \cmark & \cmark \\
\bottomrule
\end{tabular}
\vskip -10pt
\end{table}

We expand on the comparison with existing language-interfaced protein design benchmarks. Below, we revisit each benchmark in detail along the three limitations: coverage of mechanistic protein design competencies, rigor of in silico validation, and breadth of design intents under evaluation. Then we also provide a comparison against language-interfaced benchmarks focusing on protein recognition or retrieval. We summarize the distinctions between \system{} and prior benchmarks in \cref{tab:bench-comparison}.

\paragraph{Coverage of mechanistic protein design competencies.}
PDFBench~\citep{pdfbench} and InstructProBench~\citep{instructpro} restrict evaluation to function-conditioned sequence generation, leaving protein understanding and rationale-guided engineering unexamined. Mol-Instructions~\citep{Mol-instructions} and the evaluation sets released with SciReasoner~\citep{scireasoner} cover a wider set of biomolecular tasks, including protein property understanding and functional generation, but do not incorporate an engineering task. More importantly, these benchmarks do not probe whether LLMs can internalize and apply mechanistic logic as human practitioners do in real-world workflows. In practice, protein design is guided by mechanistic principles grounded in sequence--structure--function relationships, where practitioners examine structural and biochemical signals and apply targeted modifications accordingly. Consequently, they cover only a partial slice of the competencies involved in vibe protein design, and do not unify these competencies into a coherent evaluation of design as a workflow that connects understanding, rationale-guided modification, and generation.

\paragraph{Rigor of in silico validation.}
Many benchmarks rely on metrics that do not systematically verify the quality of designed proteins through in silico validation. SciReasoner~\citep{scireasoner} adopts surface-level assessment such as sequence alignment and valid amino-acid composition. Mol-Instructions~\citep{Mol-instructions} aligns each generated sequence against reference proteins in the corresponding functional regions, which conflates functional consistency with similarity to known proteins. These benchmarks do not comprehensively evaluate whether generated sequences are structurally plausible or functionally consistent at the level of folded protein outcomes.

\paragraph{Breadth of design intents.}
Finally, existing benchmarks predominantly express functional intent through semantic language-based descriptions such as Gene Ontology terms or textual annotations~\citep{pdfbench, Mol-instructions, scireasoner}, and rarely supply novel constraints such as newly characterized binding partners. They also omit practical laboratory constraints, including miniprotein length regimes and binding-site specifications, that frequently arise in therapeutic and diagnostic design. Taken together, these gaps leave open whether current LLMs can perform protein design as a coherent workflow grounded in mechanistic reasoning, verifiable biological outcomes, and realistic design constraints.

\noindent\textbf{Language-interfaced benchmark for protein understanding and retrieval.}
A complementary line of work evaluates protein QA, protein-text retrieval, and leakage-aware protein understanding benchmarks~\citep{pqa, ProteinLMBench, protein2text, protdescribe, proinstood, liveproteinbench}. These benchmarks assess whether models can answer questions about proteins or align protein sequences with textual descriptions. However, their scope is largely confined to understanding or retrieval, whereas \system{} evaluates targeted sequence modification and novel protein generation under specified functional constraints.

\section{Human Expert Review}\label{G}

\begin{figure}[H]
    \centering
    \includegraphics[width=\linewidth]{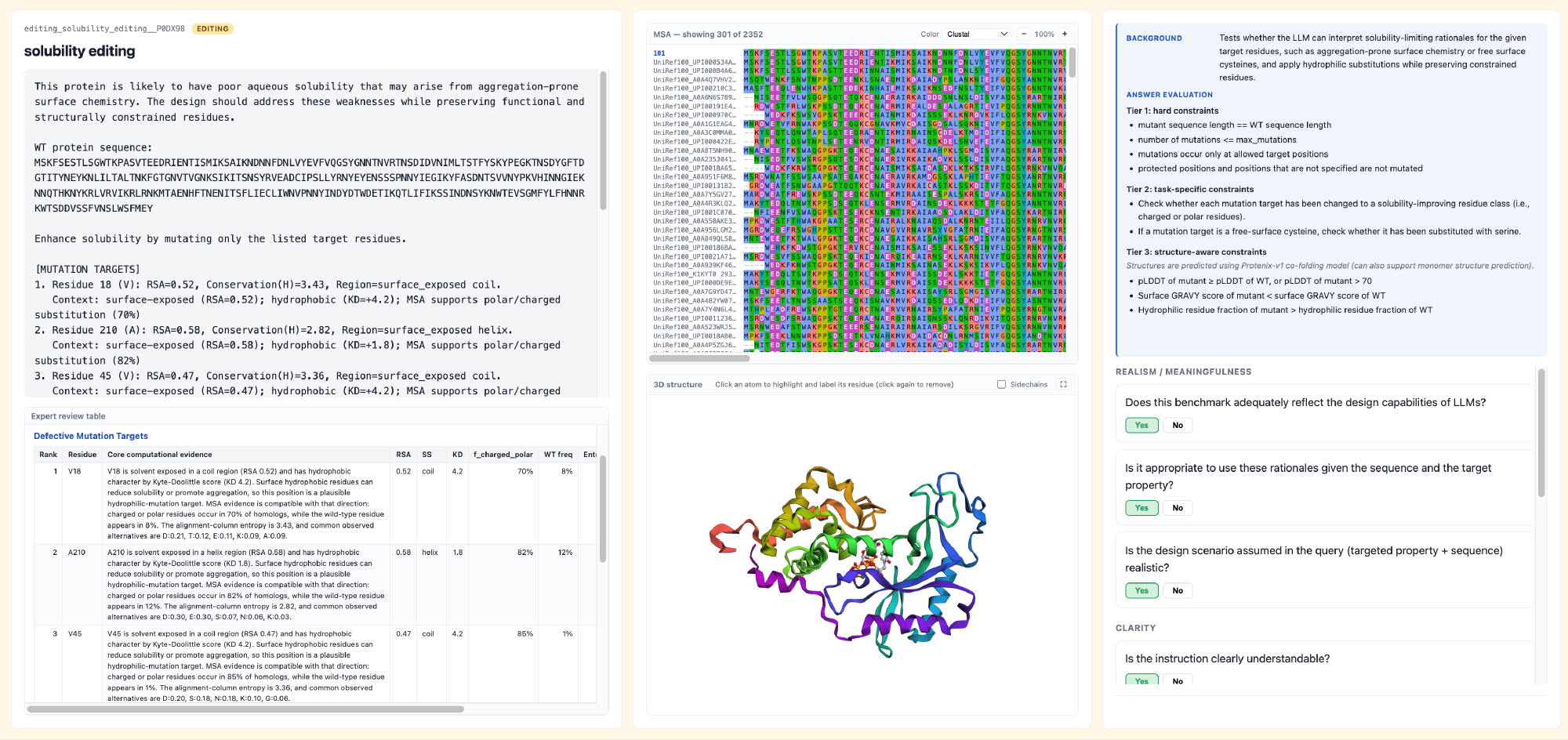}
    \caption{
    Snapshot of expert review interface for solubility engineering query.
    }
    \label{fig:expert-review-interface}
    \vskip -10pt
\end{figure}

We further validate the task construction and evaluation criteria of \system{} through review by human domain experts. Each subtask is presented to the experts together with its background, including evolutionary and structural context, the computational tools used to construct it, and the answer evaluation method. The experts then score each query against three binary rubrics: realism and meaningfulness, clarity, and in silico verifiability. The snapshot of review interface and full rubric definitions are provided in \cref{fig:expert-review-interface} and \cref{tab:expert-review-rubric}, respectively.

\begin{table}[t]
\centering
\caption{Human expert review rubrics for recognition and engineering queries. Each subtask and query is scored against three binary criteria: \emph{realism/meaningfulness}, \emph{clarity}, and \emph{in silico verifiability}. The criterion is highlighted in \textbf{bold} and the guiding questions presented to the experts are listed below.}
\label{tab:expert-review-rubric}
\small
\setlength{\tabcolsep}{6pt}
\renewcommand{\arraystretch}{1.3}
\begin{tabular}{p{0.47\textwidth} | p{0.47\textwidth}}
\hline
\multicolumn{2}{c}{\textbf{Human Expert Review Rubrics}} \\
\hline
\multicolumn{1}{c|}{\textbf{Recognition}} & \multicolumn{1}{c}{\textbf{Engineering}} \\
\hline

\textbf{Realism/meaningfulness}
\begin{itemize}[leftmargin=*, topsep=2pt, itemsep=1pt]
  \item Does this benchmark adequately reflect the design capabilities of LLMs?
  \item Does possessing this knowledge meaningfully contribute to performing design tasks?
\end{itemize}
&
\textbf{Realism/meaningfulness}
\begin{itemize}[leftmargin=*, topsep=2pt, itemsep=1pt]
  \item Does this benchmark adequately reflect the design capabilities of LLMs?
  \item Is it appropriate to use these rationales given the sequence and the target property?
  \item Is the design scenario assumed in the query (targeted property and sequence) realistic?
\end{itemize}
\\
\hline

\textbf{Clarity}
\begin{itemize}[leftmargin=*, topsep=2pt, itemsep=1pt]
  \item Is the instruction clearly understandable?
  \item Can domain experts solve this task with their own knowledge or with computational tools?
\end{itemize}
&
\textbf{Clarity}
\begin{itemize}[leftmargin=*, topsep=2pt, itemsep=1pt]
  \item Is the instruction clearly understandable?
  \item Can domain experts solve this task with their own knowledge or with computational tools?
\end{itemize}
\\
\hline

\textbf{In silico verifiability}
\begin{itemize}[leftmargin=*, topsep=2pt, itemsep=1pt]
  \item Is the ground truth unambiguous given a well-defined computational method?
\end{itemize}
&
\textbf{In silico verifiability}
\begin{itemize}[leftmargin=*, topsep=2pt, itemsep=1pt]
  \item Does the evaluation method allow for reasonably verifying the answer to the query?
\end{itemize}
\\
\hline
\end{tabular}
\vskip -10pt
\end{table}
\section{Dataset statistics}\label{H}
We report detailed benchmark statistics in \cref{tab:dataset-statistics}. The
engineering stage is smaller because its per-query content has to be validated, whereas recognition queries of the same type share one template. Although
both recognition and engineering use established computational tools, the resulting features serve
different purposes. In recognition, a computed property or curated annotation directly
defines the reference answer under a shared query template, so expert review focuses on
whether the task and its answer definition are appropriate. In engineering, these
features are used to infer which residues should be modified or protected. The rationale
defining them is authored by domain experts, but its instantiation on a given protein is
automatic, computed from structural features, ligand geometry, and Henikoff-weighted MSA
statistics. We therefore present each query to an expert together with its structure
rendering, MSA, and per-residue features, and ask whether the selected residues actually
satisfy the rationale (Appendix~\ref{G}). This check cannot be amortized across
the benchmark, and a misassigned residue can change the evaluation result for that query.
We therefore prioritize careful, expert-driven construction over sheer volume, as
comparable expert-curated benchmarks do; for example, CMT-Benchmark~\citep{cmtbenchmark} contains
50 condensed-matter-theory problems and InfiBench~\citep{infibench} contains 234 coding
questions.
\begin{table}[t]
\centering\small
\caption{Dataset statistics for \system{}.}
\label{tab:dataset-statistics}
\setlength{\tabcolsep}{4pt}
\renewcommand{\arraystretch}{0.95}
\begin{minipage}[t]{0.49\textwidth}
\centering
\begin{tabular}{l r}
\toprule
Subtasks & \# Queries \\
\midrule
\multicolumn{2}{>{\columncolor{piggypink}}l}{\emph{Recognition -- Sequence}} \\
Motif detection & 74 \\
Protein charge state classification & 57 \\
Protein hydropathy classification & 57 \\
Protein aromaticity prediction & 57 \\
Residue charge identification & 57 \\
Residue hydropathy identification & 57 \\
\emph{Subtotal} & \emph{359} \\

\multicolumn{2}{>{\columncolor{piggypink}}l}{\emph{Recognition -- Structure}} \\
Secondary structure identification & 25 \\
Burial classification & 51 \\
Disulfide bond identification & 5 \\
\emph{Subtotal} & \emph{81} \\
\multicolumn{2}{>{\columncolor{piggypink}}l}{\emph{Recognition -- Function}} \\
Family classification & 66 \\
Superfamily classification & 64 \\
GO term identification (MF) & 11 \\
GO term identification (BP) & 7 \\
GO term identification (CC) & 3 \\
Fold recognition & 36 \\
\emph{Subtotal} & \emph{187} \\
\midrule
\textbf{Recognition total} & \textbf{627} \\
\bottomrule
\end{tabular}
\end{minipage}\hfill
\begin{minipage}[t]{0.49\textwidth}
\centering
\begin{tabular}{l r}
\toprule
Subtask & \# Queries \\
\midrule
\multicolumn{2}{>{\columncolor{piggypink}}l}{\emph{Engineering}} \\
Solubility engineering & 20 \\
Stability (fold stability) engineering & 10 \\
Stability (thermostability) engineering & 10 \\
Activity (binding affinity) engineering & 9 \\
Activity (pocket extension) engineering & 11 \\
\emph{Subtotal} & \emph{60} \\
\multicolumn{2}{>{\columncolor{piggypink}}l}{\emph{Engineering -- constraint-free}} \\
Solubility engineering & 20 \\
Stability (fold stability) engineering & 10 \\
Stability (thermostability) engineering & 10 \\
Activity (binding affinity) engineering & 9 \\
Activity (pocket extension) engineering & 11 \\
\emph{Subtotal} & \emph{60} \\
\midrule
\textbf{Engineering total} & \textbf{120} \\
\midrule
\multicolumn{2}{>{\columncolor{piggypink}}l}{\emph{Generation -- Coarse-grained}} \\
GO-MF term conditioned & 30 \\
GO-MF \& GO-CC term conditioned & 30 \\
GO-MF \& GO-BP term conditioned & 30 \\
\emph{Subtotal} & \emph{90} \\
\multicolumn{2}{>{\columncolor{piggypink}}l}{\emph{Generation -- Target-specific binder}} \\
Small-molecule target & 30 \\
Protein target & 30 \\
Protein target \& miniprotein & 30 \\
Protein target \& binding-site conditioned & 30 \\
\emph{Subtotal} & \emph{120} \\
\midrule
\textbf{Generation total} & \textbf{210} \\
\midrule
\textbf{Total} & \textbf{957} \\
\bottomrule
\end{tabular}
\end{minipage}
\vskip -10pt
\end{table}

\end{document}